\documentclass[12pt]{article}

\newcommand{\beq}{\begin{equation}}
\newcommand{\eeq}{\end{equation}}
\newcommand{\beqa}{\begin{eqnarray}}
\newcommand{\eeqa}{\end{eqnarray}}
\newcommand{\beqar}{\begin{eqnarray*}}
\newcommand{\eeqar}{\end{eqnarray*}}

\newcommand{\al}{\alpha}
\newcommand{\be}{\beta}

\def\spa          {\ \ \ }
\def\non          {\nonumber}
\def\ha           {\mbox{$\frac{1}{2}$}}

\def\spa          {\ \ \ }
\def\mand         {\spa\mbox{and}\spa}

\def\Tr           {\mbox{\rm Tr}\,}
\def\STr          {\mbox{\rm STr}\,}
\def\Str          {\mbox{\rm Str}\,}

\def\cd           {{\cdot}}
\def\ran          {\rangle}
\def\lan          {\langle}
\def\fsH	{H\!\!\!\!/\,}
\newcommand{\del}{\delta}

\newcommand{\eps}{\epsilon}
\newcommand{\ga}{\gamma}
\newcommand{\Ga}{\Gamma}

\newcommand{\inn}{\!\cdot\!}

\newcommand{\lam}{\lambda}

\newcommand{\z}{\zeta}

\newcommand{\ie}{{\it i.e.,}\ }
\newcommand{\labell}[1]{\label{#1}} 
\newcommand{\reef}[1]{(\ref{#1})}
\newcommand\prt{\partial}
\newcommand\veps{\varepsilon}

\newcommand\cL{{\cal L}}
\newcommand\cD{{\cal D}}

\newcommand\bz{\bar{z}}

\newcommand\cT{T}

\begin{document}
\baselineskip 18pt%
\begin{titlepage}
\vspace*{.89mm}%
\hfill
\vbox{

    \halign{#\hfil         \cr
         ICTP-PH-TH/2012-xyz\cr
           } 
      }  
\vspace*{8mm}
\vspace*{5mm}%
\center{ {\bf \Large On higher derivative corrections to Wess-Zumino
and Tachyonic actions in type II super string theory
 }}\vspace*{3mm} \centerline{{\Large {\bf  }}}
\vspace*{3mm}
\begin{center}
{Ehsan Hatefi}$\footnote{
E-mail:ehatefi@ictp.it}$

\vspace*{0.8cm}{ {
International Centre for Theoretical
Physics, Strada Costiera 11, Trieste, Italy}}

\vspace*{.75cm}
\end{center}
\begin{center}{\bf Abstract}\end{center}
\begin{quote}
We evaluate in detail the string scattering amplitude to compute different
  interactions of two massless
scalars, one tachyon and one closed string Ramond-Ramond field in type II super string theory. In particular we find two scalar field and two tachyon couplings to all orders of $\alpha'$ up to on-shell ambiguity. We then obtain the momentum expansion of this amplitude and apply this  infinite number of couplings to actually check that the infinite number of  tachyon poles of  S-matrix element of this amplitude for the $p=n$
case (where $p$ is the spatial dimension of a D$_p$-brane  and $n$ is the rank of a Ramond-Ramond field strength ) to all orders of $\alpha'$ is precisely equal  to the infinite number of   tachyon poles of the field theory. In addition to confirming the couplings of closed string Ramond-Ramond field  to the world-volume gauge field and
scalar fields  including commutators,
we also
propose an extension of the Wess-Zumino action
 which  naturally
reproduces these new couplings in field theory such that they could be confirmed with direct S-matrix computations.
Finally we show that the infinite number of massless poles and contact terms of this amplitude for the $p=n+1$ case can be reproduced by Chern-Simons, higher derivative corrections of the Wess-Zumino and symmetrized trace tachyon DBI
actions.
\end{quote}
\end{titlepage}
\section{Introduction}
 D$_p$-branes  must be regarded as the sources of Ramond-Ramond $(p+1)$-form fields in type II super string theories \cite{Polchinski:1995mt}. Their perturbative excitations  should be understood as  fundamental open string states on their
world volume.
More details can be found in \cite{Witten:1995im,Polchinski:1996na}. Having taken into account N coincident D-branes, we must have  U(N) non abelian symmetry \cite{Witten:1995im}. Notice that the bosonic action was derived in \cite{Myers:1999ps,Taylor:1999pr}.
\vskip 0.1in

The low energy action  representing  D-branes' dynamics defined as Born-Infeld action :
\beq
{S}_{BI}=-T_p \int d^{p+1}\sigma\,\STr\left(e^{-\phi}\sqrt{-\det\left(
P\left[E_{ab}+E_{ai}(Q^{-1}-\delta)^{ij}E_{jb}\right]+
\lambda\,F_{ab}\right)\,\det(Q^i{}_j)}
\right),
\labell{finalbi}
\eeq
with
\beq
E_{ab}=G_{ab}+B_{ab}
\qquad{\rm ,}\qquad
Q^i{}_j\equiv\delta^i{}_j+i\lambda\,[\Phi^i,\Phi^k]\,E_{kj},
\labell{extra6}
\eeq
 For more details see \cite{Myers:1999ps,Hatefi:2010ik}. Remember different couplings in this action  for BPS branes have been shown to be consistent with  disk level amplitudes in  super string
theory \cite{Hatefi:2010ik,Hatefi:2012ve,Hatefi:2012rx,Hatefi:2011jq}. The important point should be made is that
derivatives of the  field strength of the gauge field, and the second and higher derivatives of
the scalars  must be embedded to this action(see \cite{ Hatefi:2012ve,Hatefi:2012rx}).

 We should introduce the Wess-Zumino action as well. This action in the presence of closed string RR field (showed by $C$)makes sense as follows \cite{Polchinski:1995mt,li1996a}

\beq
S_{WZ}=\mu_p\int \STr\left(P\left[e^{i\lambda\,i_\Phi i_\Phi} (
\sum C^{(n)})\right]
e^{\lambda\,F}\right)\ .
\labell{finalcs}
\eeq
Note that we have set
\beq
 G_{\mu\nu}=\eta_{\mu\nu} ,\quad  B_{\mu\nu}=\Phi=0
\nonumber\eeq
 so one may believe that we are working on flat space background. In our conventions  $\lambda=2\pi\alpha'$.
We want to begin with non-supersymmetric D-branes in type II
theory where $p$ is odd for IIA and even for IIB. In the spectrum of a  non-BPS D-brane in type II theory, there exists a tachyonic state (indicated by $T$), a massless gauge field (showed by $A_a$) and some massless scalars $\phi^i$ and some fermions (for example see \cite{Sen:1999mg}).  All scalars might express transverse
oscillations of the brane.

Now the open string tachyon must be condensed to a kink to produce a stable $D_{p-1}$-brane. To study brane production in detail see
~\cite{Horava:1998jy,Bergman:1998xv,Witten:1998cd}. It is worth  mentioning the point that  unstable branes first have been used in checking some duality conjectures~\cite{Sen:1998rg}.
\\

Based on some concrete points \cite{Sen:2004nf}, one might expect that the effective theory of non-BPS branes should involve   tachyon and   massless states within itself. Therefore the action for   non-BPS branes  is constructed from two  parts such that
\beqa
S_{non-BPS}&=&S_{DBI}+S_{WZ}\labell{nonbps}\nonumber\eeqa

Now tachyon must get involved in Dirac-Born-Infeld and Wess-Zumino effective actions.
Having done boundary string field theory (BSFT) method, we may talk about these effective actions, see \cite{Kraus:2000nj,Takayanagi:2000rz} for further details.
Due to some efforts \cite{Kraus:2000nj,Takayanagi:2000rz,Kennedy:1999nn} tachyons have already  been taken into account in the WZ effective action using the superconnection of noncommutative geometry \cite{Roepstorff:1998vh}. Concerning super connection~\cite{Roepstorff:1998vh}, one can write down the WZ action as follows
 \beqa
S_{WZ}&=&\mu'_p \int_{\Sigma_{(p+1)}} C \wedge \STr e^{i2\pi\alpha'\cal F}\labell{WZ56},\eeqa
where $\Sigma_{(p+1)}$ is the world volume, $\mu'_p$ is the RR charge of branes. The curvature of super connection first has been derived for brane anti brane system \cite{Garousi:2007fk} then the method was extended to include non-BPS branes  \cite{Garousi:2008tn} and it is obtained as

 \begin{displaymath}
i{\cal F} = \left(
\begin{array}{cc}
iF -\beta'^2 T^2 & \beta' DT \\
\beta' DT & iF -\beta'^2T^2
\end{array}
\right) \ ,
\non\end{displaymath}
One must emphasize that superconnection's structure in the WZ action first has been achieved with the direct S-matrix method as appeared in \cite{Kennedy:1999nn}.
The tensor structure of this object is such that the definitions of field strength of the gauge field and covariant derivative of tachyon  are
 \beqa
 F=\frac{1}{2}F_{ab}dx^{a}\wedge dx^{b}\quad, DT=(\partial_a T-i[A_{a},T])dx^{a}
 \eeqa

 where $\beta'$ is a  normalization tachyon which is constant with dimension $(1/\sqrt{\alpha'})${\footnote{$\alpha'=l_s^2$ and $l_s$ becomes string length scale.}.
 It is shown that
$\beta'=\frac{1}{\pi} \sqrt{\frac{6\ln(2)}{\alpha'}}\labell{norm0}$ \cite{Garousi:2008ge}. More consistency of the WZ action with other S-matrix elements can be seen in \cite{Garousi:2007fk,Garousi:2008ge}.
Applying the expansion for the exponential term in the WZ action \reef{WZ56}, one gets \cite{Garousi:2008tn}
\beqa
\mu_p'(2\pi\alpha')C\wedge \Str i{\cal F}&\!\!\!\!=\!\!\!&2\beta'\mu_p' (2\pi\alpha')\Tr\left(C_{p}\wedge DT\right),\labell{exp2}\\
\frac{\mu_p'}{2!}(2\pi\alpha')^2C\wedge \Str i{\cal F}\wedge i{\cal F}&\!\!\!\!=\!\!\!\!&2\beta'\mu_p'(2\pi\alpha')^2\Tr\left(C_{p-2}\wedge DT\wedge F\right).
\nonumber
\eeqa

To study  the effective action of non-BPS branes we would like to proceed with the second approach, which is the S-matrix formalism. In this method one must include  kinetic term of the tachyon  in the DBI part  \cite{Garousi:2000tr}; however, due to considering internal degrees of freedom or internal Chan-Paton factors belonging to non-BPS branes this action gets modified \cite{Garousi:2008ge,Garousi:2008nj}.

$\frac{}{}$

Morever, according to the S-matrix method, one subtlety  around the unstable point of the tachyon DBI does exist.
To be more specific, all massless fields must carry an identity  internal CP matrix, while tachyon in the (0)-picture  has to carry   $\sigma_1$ and  it does carry the CP matrix of $\sigma_2$ in the (-1)-picture. Note also that the picture changing operator carries $\sigma_3$ internal  CP matrix \cite{DeSmet:2000je}. Notice that, this important point has not been confirmed yet  around the stable point of the tachyon action.
$\frac{}{}$

\vskip 0.1in

The outline of the paper is as  follows. In section 3 we are going to compute in detail a tree-level physically four point (technically five point) string scattering amplitude including one RR, two scalar field and one tachyon vertex operators in the world volume of  type II super string theory. The world-volume theory might be
 rewritten  in terms of  an infinite number of the derivatives of the fields as one of the main goals of the paper is indeed obtaining those infinite numbers of the couplings between two tachyons and two scalar fields  to all orders of $\alpha'$. Keeping in mind that once  the world volume fields vary so slowly, we must get the usual effective theory by reducing the higher derivative theories.

 $\frac{}{}$

In section 4 we carry out the momentum expansion whose leading order terms must be consistent with  field theory. On the other hand, the rest of the terms which are non leading terms must correspond to the higher derivative corrections of tachyon DBI and Wess-Zumino effective  actions. We conjecture that we found a unique momentum expansion for all four point functions including one RR, one tachyon and two massless open strings which can be either two gauge fields or two scalar  fields. We guess this happens also in the amplitude of one RR, one tachyon, one gauge field and one scalar field but it has not been checked yet so we postpone it for future work  \cite{Hatefi:2012eh1}. In section 5 we talk about effective field theory on the world volume of brane. In section 5.1 using symmetrized trace tachyon DBI  action we reproduce the first tachyon pole in field theory  for $p+1=n$ case,
  where $p$ is the spatial dimension of a D$_p$-brane  and $n$ is the rank of a RR field strength.

 \vskip 0.1in

 To obtain the infinite number of tachyon
  poles for this case, one needs to know the higher derivative couplings of two scalar fields and two tachyons up to all orders of $\alpha'$. Applying the string computations and making use of some how T-duality transformation in section 6 we find higher derivative couplings of two scalar fields and two tachyons up to all orders of $\alpha'$  within on-shell ambiguity. Then using the infinite number of the couplings obtained in field theory we will show that the  infinite number of tachyon poles of this S-matrix element are exactly reproduced to all orders of $\alpha'$. In section 7.1 we obtain all infinite numbers of massless poles in field theory for $p=n+1$ case.
In order to get consistent results between string theory and field theory we propose an extension of the Wess-Zumino
action and discover some sort of new couplings which could be confirmed by direct S-matrix computations.

\vskip 0.2in

We also  generate  all infinite contact terms of this amplitude. Eventually we end up with a discussion, mention our results, and give some hints  for future directions. Appendix A  consists of some information about doubling trick in II super string theory, some useful comments
on conformal field theory propagators, and
some correlation functions including two spin operators and some number of fermion fields and/or currents. Appendix B  includes some useful integrals for five point functions. Let us address our notation.

$\mu,\nu$ represent the entire   space-time dimensions and $\mu,\nu= 0,...,9 $ and $a, b, c$ indices show world-volume space $a, b, c  = 0, 1,..., p$ and finally $i,j$ indices indicate
   transverse space,$i,j  = p + 1,...,9$.

\section{Remarks on Scattering amplitude}

The S-Matrix technique is a very important tool in super string theory to actually discover new couplings between mixed combinations of open and closed
strings. Therefore corrections in field theory might have been obtained in $\alpha'$ by standard scattering amplitude methods.
To describe  scattering of closed strings with open strings see \cite{Hatefi:2010ik,Polchinski:1994fq}.
To observe one of the important aspects of D-brane physics we refer the reader to  \cite{Park:2007mc
}.
In order to follow the scattering argument  and its applications, it is worth looking at\cite{Hashimoto:1996bf}.

\subsection{Motivations}
Pursuing scattering amplitudes has  been one of the most  active fields during the recent years \cite{Britto:2004ap}.
The (Britto,Cachazo,Feng and Witten) BCFW recursion relations were derived by taking into account the fact that an $n$-point tree level amplitude is expressed  in terms of the rational function of all external momenta. Thus by applying  analytic continuation of all momenta to an entire complex plane, the amplitude would be constructed by  its poles.
Some more attempts have been made in \cite{ArkaniHamed:2008yf,Cheung:2008dn} which help  one to realize which kinds of theories should be determined by  the BCFW  relations.

The  string BCFW relations have been investigated in \cite{Cheung:2010vn,Fotopoulos:2010cm} for the scattering amplitude of external open string tachyons.

Because of  some kinematic reasons it is not possible to embed their efforts within our formalism which is perturbative string calculations. We  will notice these arguments in the section of momentum expansion. In fact
our computations make sense just in the presence of a constant value of RR 's momentum which is indeed $p_ap^a\rightarrow\frac{1}{4}$, and it is going to be the key point in our calculations.
\\

The second motivation for studying unstable objects  perhaps is
 realizing properties of string theory in time-dependent
backgrounds \cite{Gutperle:2002ai,Sen:2002nu}.
Open string tachyons are showing us the instability of the processes, which we are interested in. A. Sen in \cite{Sen:1999md}
has shown that tachyon DBI action \cite{Bergshoeff:2000dq}
 might be able to represent decay  non-BPS D-branes
\cite{Sen:2002an} around the stable point of  tachyon potential.

\vskip 0.1in

Another motivation to follow unstable branes is in fact  studying  spontaneous chiral symmetry breaking  in  some holographic patterns in QCD \cite{Casero:2007ae}.
In order to have a formalism for inflation (of course in string theory), one may use  D-brane anti D-brane effective action\cite{Garousi:2007fk} gets divided along an extra dimension \cite{Dvali:1998pa}.
In this formalism the brane separation has to play  inflaton's role. When this distance is smaller than the string scale, the open strings stretching between them will become two real tachyon  modes and they may condense \cite{Banks:1995ch,Sen:1998ii,deAlwis:1999cg}. Indeed the negative
energy density of the tachyon field which is condensed  cancels  the
positive energy density of the brane anti brane   and
gives us a $D_{p-2}$-brane with a finite tension~\cite{Sen:1998sm}. Then inflation ends and finally the energy of inflaton
will decay  to the particles in the Standard model \cite{Kofman:1997yn}.
To follow some of the applications of this action in cosmology references  \cite{Alexander:2001ks} may be worth referencing.

\vskip 0.1in

Having read  the boundary conformal field
theory (BCFT), some of the results such as producing a pressure less gas with nonvanishing
energy density at the end of  tachyon
condensation \cite{Sen:2002in} may be obtained from this action at the minimum of the tachyon potential.
 Although the higher derivative terms might have a significant
role at the top of the tachyon potential, they  have small
effects at the minimum of the potential. Note that the higher derivative terms
 are not involved in the BSFT action. Thus, the action which was obtained through
 S-matrix elements at the top of
the potential may indeed represent the effective action of string theory at the
minimum of the tachyon potential.
 Notice that the action is also consistent with T-duality rules.

On the other hand, the effective action for brane-antibrane \cite{Hatefi:2012eh3} based on the S-matrix elements computations is found in \cite{Garousi:2007fk} to be

\beqa
S_{DBI}&=&-T_p\int
d^{p+1}\sigma \STr\left(V({\cal T})
\sqrt{-\det(\eta_{ab}
+2\pi\alpha'F_{ab}+2\pi\alpha'D_a{\cal T}D_b{\cal T})} \right),\labell{nonab} \eeqa

\vskip 0.1in

 STr means symmetric trace for all matrices including
 $F_{ab},D_a{\cal T}$  and
${\cal T}$ in the potential. To see more information and the explicit form of  these matrices  see \cite{Garousi:2007fk}.
The tachyon potential that does make an exact result with S-matrix computations
has an expansion as:
\beqa
V(|T|)&=&1+\pi\alpha'm^2|T|^2+
\frac{1}{2}(\pi\alpha'm^2|T|^2)^2+\cdots
\label{tpot}\eeqa
which is consistent expansion in comparing with the tachyon potential in BSFT which has $V(|T|)=e^{\pi\alpha'm^2|T|^2}$ potential form
\cite{Kutasov:2000aq}.
It is also consistent with the sigma model effective action \cite{Tseytlin:2000mt}.
Note that $m^2$ is  tachyon's mass square and
$m^2=-1/(2\alpha')$.

\vskip 0.1in

Having expanded the square root,
and renormalized   $\frac{1}{2}\pi T^2\rightarrow 2T^2$,
we get
\beqa
S=-T_p\,e^{-T^2}\left(1+4\alpha'D_aTD^aT+\cdots\right)\nonumber\eeqa
which is precisely the term coming from  BSFT action suggested in
\cite{Minahan:2000tf} .

\vskip 0.1in

Therefore using the S-matrix method, we are able to find out either  tachyon action around the unstable point of non-BPS D-branes or D-brane anti D-brane where the higher derivatives of  tachyon are indeed important. So not only  is it really interesting to study them but also worth it to obtain their general form  to all orders of $\alpha'$.

\vskip 0.1in

 The existence of  the coupling $\Tr(C_{p-2}\wedge F\wedge DT)$ has already been checked in \cite{Garousi:2008ge} by working out in detail the disk level S-matrix element of one RR field, one tachyon and two gauge fields in the world volume of a single non-BPS  brane. In this  paper  among other things, we find some new contact interactions of the form $\Tr(C_{p-2}\wedge DT\wedge D\phi^i\wedge D\phi_i)$   $\frac{}{}$,
  $\Tr(C_{p}\wedge DT \phi^i\phi_i) $, $
 \Tr(\partial_{a_0}T[\Phi^i,\Phi^j])C^{(p+2)}_{jia_1\cdots a_p}(\veps^v)^{a_0\cdots a_p}$ and some other couplings which can be confirmed just by S-matrix computations  and fix their coefficients using the  S-matrix elements of one RR field, one tachyon  and two scalar fields.

\section{The Scattering  amplitude between one closed string RR, two scalar fields and one tachyon
}
By making use of the conformal field theory techniques, we might carry out the string scattering amplitude to find out all couplings of one closed string RR field in the bulk  to two open string scalar fields and one open string  tachyon on the world-volume of a single non-BPS D-brane with flat empty space background. To compute a S-matrix element, one must fix the picture of the vertex operators .

\vskip 0.1in

Knowing the fact that
the total super ghost charge for disk level amplitude must be -2 , we will choose the vertex operators accordingly.
Concerning string duality, we map the disk to the upper half plane thus the boundary of the disk
becomes the real axis. Therefore, the closed string vertex operator should be inserted at the middle and all open string vertex
operators must be put at the boundary of the disk world-sheet. It is fair to say that
 some  efforts in order for obtaining the string scattering amplitudes at tree level in both BPS and non BPS formalism have been made
\cite{Hatefi:2010ik,Hatefi:2012ve,Hatefi:2012rx,Hatefi:2011jq,Kennedy:1999nn,Garousi:2007fk, Garousi:2008ge,Stieberger:2009hq}.
The external states will be appeared in our amplitude as the following:
\beqa
{\rm tachyon}&& T,\ k_3,
\nonumber\\
{\rm transverse\ scalars:}&& \Phi_1^i,\  k_1,
\nonumber\\
&&\Phi_2^j,\ k_2,
\nonumber\\
{\rm RR}\ (p+1){\rm -form:}&& C^{p+1},\  p\,\,\, .
\labell{states}\nonumber
\eeqa
We also define
\beqa
s&=&-\frac{\alpha'}{2}(k_1+k_3)^2,\qquad t=-\frac{\alpha'}{2}(k_1+k_2)^2,\qquad u=-\frac{\alpha'}{2}(k_2+k_3)^2.
\eeqa

Applying momentum conservation along the world volume of the brane, we get
  the following relation
   \beqa
   s+t+u=- p^a p_a -\frac{1}{4}\label{momencon}.\eeqa
Remember the fact that the vertex operators of a  non-BPS D-brane must carry internal degrees of freedom or the internal Chan-Paton(CP) matrix \cite{Sen:1999mg}.
In the other words in order to distinguish the form of closed string vertex operators of non BPS branes from their form in the brane-antibrane system an internal CP matrix is needed. By setting tachyon  to zero, we may conclude that both the effective field theory of brane-antibrane and non-BPS branes must be  reduced to  effective field theory of just BPS branes. Therefore we come to the key point  which is imposing an identity internal  CP matrix for all massless fields in both brane-antibrane and non-BPS branes' formalism.
\vskip 0.1in

As an  example, the RR field in effective theory of brane-antibrane does include an identity matrix. This is related to the fact that by setting tachyon to zero the Wess-Zumino action of brane-antibrane  does go back to the Wess-Zumino action of two BPS branes. We devote this identity internal CP matrix to RR, gauge field  and scalar field vertex operators in the $(0)$-picture. In \cite{DeSmet:2000je} it has been shown that the picture changing operator does carry internal CP matrix $\sigma_3$ and  so  in the (-1)-picture, the internal CP matrix of RR vertex operator in the brane-antibrane system and gauge and scalar (in both non-BPS brane and brane-antibrane system)  is not identity any more and in fact it is $\sigma_3$.

\vskip 0.2in

Notice that in non-BPS branes, there must be  an extra factor of $\sigma_1$ in  the RR vertex operator \cite{Sen:1999mg,Garousi:2008tn},
  so the RR vertex operator of  a non-BPS brane in the $(0)$-picture must involve the internal CP matrix $\sigma_1$. We are going to address this CP matrix to tachyon vertex operator in  $(0)$-picture as well. Therefore by applying the picture changing operator to RR in the $(0)$-picture we reach the point that RR vertex operator in the (-1)-picture in the non-BPS formalism has to carry the internal matrix $\sigma_3\sigma_1$.
Finally by the same argument we conclude that the internal CP matrix of
the tachyon  in the (-1)-picture is $\sigma_2$.

\vskip 0.1in

Thus, the S-matrix element of one RR field, two scalar fields and one tachyon
in the world volume of a single non-BPS D-brane   in super string theory may be given with :
\begin{eqnarray}
{\cal A}^{\phi\phi TC} & \sim & \sum_{\rm non-cyclic}\int dx_{1}dx_{2}dx_{3}dzd\bar{z}\,
  \lan V_{\phi}^{(0)}{(x_{1})}
V_{\phi}^{(0)}{(x_{2})}V_T^{(0)}{(x_{3})}
V_{RR}^{(-2)}(z,\bar{z})\ran,\labell{sfield}\eeqa

The internal  CP factor is $\Tr(\sigma_1II\sigma_1)=2 $ for all permutations of the scalar fields. Note that in the non-BPS system both  the Ramond-Ramond and tachyon in the (-2)-picture do carry $\sigma_1$ internal CP matrix and this is related to the point that the amplitude of $CTA$ makes sense in the world volume of non-BPS branes. However by taking into account the amplitude of $CTT$ in the world volume of brane-antibrane one comes over to the CP factor of RR in the (-2)-picture as the identity CP matrix in this system. It is indeed so easy to compute this amplitude by putting the RR and tachyon in the(-1)-picture so that

\begin{eqnarray}
{\cal A}^{\phi\phi TC} & \sim & \sum_{\rm non-cyclic}\int dx_{1}dx_{2}dx_{3}dzd\bar{z}\,
  \lan V_{\phi}^{(0)}{(x_{1})}
V_{\phi}^{(0)}{(x_{2})}V_T^{(-1)}{(x_{3})}
V_{RR}^{(-1)}(z,\bar{z})\ran.\labell{sstring}\eeqa
The  CP factor now becomes  $\Tr(\sigma_3\sigma_1II\sigma_2)=2i$ for 123 and 132 orderings. Thus to investigate the S-matrix element \reef{sfield}, which is much more difficult than \reef{sstring}, we do computations of the amplitude of  \reef{sstring} and finally multiply  a coefficient of $(-i)$ in the final result.
\\

In general the form of vertex operators in \reef{sstring} is written down  as \footnote{In string side, we used to set  $\alpha'=2$.}
\beqa
V_{T}^{(-1)}(y) &=& e^{-\phi(y)} e^{2ik\cd X(y)}\lam\otimes\sigma_2,
\nonumber\\
V_{\phi}^{(0)}(x) &=& \xi_{i}\bigg(\partial X^i(x)+2iq\cd\psi\psi^i(x)\bigg)e^{2iq.X(x)}\lam\otimes I,
\nonumber\\
V_{RR}^{(-1)}(z,\bar{z})&=&(P_{-}\fsH_{(n)}M_p)^{\al\be}e^{-\phi(z)/2} S_{\al}(z)e^{ip\cd X(z)}e^{-\phi(\bar{z})/2} S_{\be}(\bar{z}) e^{ip\cd D \cd X(\bar{z})}\otimes\sigma_3\sigma_1,\nonumber\\
V_\phi^{-2}(x)&=&e^{-2\phi}V_\phi^{0}(x),\nonumber\eeqa

\vskip 0.1in

where $k,q,p$ are the momenta of tachyon, scalar field and closed string RR field accordingly. Notice that the momentum of open strings has to be constrained to be inside of the world volume. The on-shell condition for the tachyon is $k^2=\frac{1}{2\alpha'}$ and for the RR and scalar it is $p^2=q^2=0$. The physical state condition for the massless scalar and RR is $k.\xi=0$ and $p_i^{\mu}\epsilon_{i\mu\mu_3...\mu_n}=0$. Also $\lam$ must be regarded as an external CP matrix which should be in the $U(N)$ gauge group.
The definition of projection operator  is $P_{-} = \ha (1-\ga^{11})$. Throughout of the paper we  work with
 the full $32\times 32$ Dirac matrices in ten dimensions of space-time.
The definition of the RR field strength  is
\begin{displaymath}
\fsH_{(n)} = \frac{a
_n}{n!}H_{\mu_{1}\ldots\mu_{n}}\ga^{\mu_{1}}\ldots
\ga^{\mu_{n}}
\ ,
\non\end{displaymath}

with $n=2,4$ for type IIA and $n=1,3,5$ for type IIB. $a_n=i$ for IIA and $a_n=1$ for IIB theory. The spin
indices must be raised with the charge conjugation matrix such that

\beqa
(P_{-}\fsH_{(n)})^{\al\be} =
C^{\al\del}(P_{-}\fsH_{(n)})_{\del}{}^{\be}
\nonumber\eeqa
 In order to deal with standard holomorphic conformal field theory propagators on the boundary of world sheet we might use the doubling trick
 (see Appendix A for more details). Implementing this  trick, we are allowed to use just the  standard  correlators for the world-sheet fields $X^{\mu},\psi^{\mu}, \phi$ as follows
\begin{eqnarray}
\lan X^{\mu}(z)X^{\nu}(w)\ran & = & -\eta^{\mu\nu}\log(z-w) , \non \\
\lan \psi^{\mu}(z)\psi^{\nu}(w) \ran & = & -\eta^{\mu\nu}(z-w)^{-1} \ ,\non \\
\lan\phi(z)\phi(w)\ran & = & -\log(z-w) .\
\eeqa

 We refer the interested reader to see  Appendix A of the paper for  applying the doubling trick, for working out with the standard holomorphic  correlators, and for finding various correlation functions including spin operators, fermions and currents .

Note also that to simplify our computations we introduce $x_{4}\equiv\ z=x+iy$ and $x_{5}\equiv\bz=x-iy$, thus  the  amplitude  for 123 ordering will be written  as
\beqa {\cal A}^{\phi\phi TC}&\sim& \int
 dx_{1}dx_{2}dx_{3}dx_{4} dx_{5}\,
(P_{-}\fsH_{(n)}M_p)^{\al\be}\xi_{1i}\xi_{2j}x_{45}^{-1/4}(x_{34}x_{35})^{-1/2}(I_1+I_2+I_3+I_4)\nonumber\\&&
\times\Tr(\lam_1\lam_2\lam_3)\Tr(II\sigma_2\sigma_3\sigma_1),\labell{125}\eeqa

with $x_{ij}=x_i-x_j$ and

\beqar I_1&=&{<:\partial X^i(x_1)e^{2ik_1.X(x_1)}:\partial X^j(x_2)e^{2ik_2.X(x_2)}
:e^{2ik_3.X(x_3)}:e^{ip.X(x_4)}:e^{ip.D.X(x_5)}:>}
 \  \non \\&&\times{<:S_{\al}(x_4):S_{\be}(x_5):>},\nonumber\\
I_2&=&{<:\partial X^i(x_1)e^{2ik_1.X(x_1)}:e^{2ik_2.X(x_2)}
:e^{2ik_3.X(x_3)}:e^{ip.X(x_4)}:e^{ip.D.X(x_5)}:>}
 \  \non \\&&\times{<:S_{\al}(x_4):S_{\be}(x_5):2ik_{2b}\psi^b\psi^j(x_2):>},\nonumber\\
 I_3&=&{<: e^{2ik_1.X(x_1)}:\partial X^j(x_2)e^{2ik_2.X(x_2)}
:e^{2ik_3.X(x_3)}:e^{ip.X(x_4)}:e^{ip.D.X(x_5)}:>}
 \  \non \\&&\times{<:S_{\al}(x_4):S_{\be}(x_5):2ik_{1a}\psi^a\psi^i(x_1):>},\nonumber\\
 I_4&=&{<: e^{2ik_1.X(x_1)}:e^{2ik_2.X(x_2)}
:e^{2ik_3.X(x_3)}:e^{ip.X(x_4)}:e^{ip.D.X(x_5)}:>}
 \  \non \\&&\times{<:S_{\al}(x_4):S_{\be}(x_5):2ik_{1a}\psi^a\psi^i(x_1):2ik_{2b}\psi^b\psi^j(x_2):>}.\eeqar

Concerning the  correlators corresponding to bosonic fields in  Appendix A, one could compute all correlators of $X$. To find out the correlator of world sheet fermions ($\psi s$) with two spin operators (coming from RR sector), we may use  Wick-like rule~\cite{Garousi:2007fk,Liu:2001qa} (see also Appendix A).

Wick-like rule might be also generalised  to obtain the correlation function of two spin operators and a number of  currents and fermion fields \cite{Hatefi:2010ik}.\vskip 0.1in

 There are two  subtleties in applying the formula \reef{wicklike} for currents. The first one  is that one should not take into account  the Wick-like contraction for two fermion fields within one  current. The second one is that one has to pay attention to all minus signs which are coming from fermion propagators once we want to write down Wick-like contraction of two fermion fields in which they belong to two different currents \cite{Hatefi:2010ik}.
 This point is playing the key role in all straightforward but tedious computations.  Considering those issues, one gets the old results
 in the presence and absense of current as they have already been checked in \cite{Garousi:2008ge}.

 \vskip 0.1in

When there are two currents, the formula \reef{wicklike} gives  a more complicated result. However,
 here there is no correlation between the transverse fields and world volume fields so we end up some how with simple result as follows
\beqa
I_4'&=&<:S_{\al}(x_4):S_{\be}(x_5):\psi^a\psi^i(x_1):\psi^b\psi^j(x_2):>\labell{63}\\
&=&\frac{1}{4}x_{45}^{3/4}(x_{14}x_{15}x_{24}x_{25})^{-1} \bigg\{(\Gamma^{jbia}C^{-1})_{\alpha\beta}+2\frac{Re[x_{14}x_{25}]}{x_{12}x_{45}}\bigg[\eta^{ab}(\Gamma^{ji}C^{-1})_{\alpha\beta}+\eta^{ij}(\Gamma^{ba}C^{-1})_{\alpha\beta}\bigg]\nonumber\\&&+4\bigg(\frac{Re[x_{14}x_{25}]}{x_{12}x_{45}}\bigg)^{2}
(-\eta^{ab}\eta^{ij})C^{-1}_{\alpha\beta}\bigg\},\nonumber\eeqa
Having performed the  $X$ correlators  and replacing \reef{63} in \reef{125}, one  may write the amplitude as
\beqa
{\cal A}^{\phi\phi TC}&\!\!\!\!\sim\!\!\!\!\!&\int dx_{1}dx_{2} dx_{3}dx_{4}dx_{5}(P_{-}\fsH_{(n)}M_p)^{\al\be}I\xi_{1i}\xi_{2j}x_{45}^{-1/4}x_{34}^{-1/2}x_{35}^{-1/2}\nonumber\\&&\times
\bigg(x_{45}^{-5/4}C^{-1}_{\al\be}(-\eta^{ij}x_{12}^{-2}+a^i_1a^j_2)+a^i_1(a^j_3)_{\al\be}+a^j_2(a^i_4)_{\al\be}-4k_{1a}k_{2b}I_4'\bigg),\labell{amp3}\eeqa
where  $I_4'$ is given in \reef{63}  and
\beqa
I&=&|x_{12}|^{4k_1.k_2}|x_{13}|^{4k_1.k_3}|x_{14}x_{15}|^{2k_1.p}|x_{23}|^{4k_2.k_3}|x_{24}x_{25}|^{2k_2.p}
|x_{34}x_{35}|^{2k_3.p}|x_{45}|^{p.D.p},\nonumber\\
a^i_1&=&-ip^{i}\bigg(\frac{x_{45}}{x_{14}x_{15}}\bigg),\nonumber\\
a^j_2&=&-ip^{j}\bigg(\frac{x_{45}}{x_{24}x_{25}}\bigg),\nonumber\\
(a^j_3)_{\al\be}&=&ik_{2b}x_{45}^{-1/4}(\Gamma^{jb}C^{-1})_{\al\be}
(x_{24}x_{25})^{-1},\nonumber\\
(a^i_4)_{\al\be}&=&ik_{1a}x_{45}^{-1/4}(\Gamma^{ia}C^{-1})_{\al\be}
(x_{14}x_{15})^{-1}.\eeqa
 As the first check of our computations, the amplitude is now invariant under
SL(2,R) transformations. One may try to cancel the  volume of the
conformal Killing group  by fixing the positions of the three open strings.
So to set gauge fixing of this  symmetry we used to fix the position of the open strings  as  \beqar
 x_{1}&=&0 ,\qquad x_{2}=1,\qquad x_{3}\rightarrow \infty,
 \qquad dx_1dx_2dx_3\rightarrow x_3^{2}
 \eeqar
 Note that different fixing for position of open strings will give rise to different ordering  in the
boundary of the world-sheet. One might consider all noncyclic
permutations of the vertices to get the correct scattering
amplitude. However, for our purpose which is comparing string theory S-matrix
elements with field theory S-matrix elements, it is really sufficient to
consider the S-matrix element with just  the factor of $\Tr(\lambda^1\lambda^2\lambda^3)$.
After  fixing the $SL(2,R)$, one ends up with  one double integral
which can be performed.
Now the integrals  are given in terms of the three Mandelstam variables.
For the integrals see Appendix B of the paper.
Applying these integrals and making use of several identities  one can eventually  read off the amplitude  \reef{amp3}  as the following :

 \beqa {\cal A}^{\phi\phi TC}&=&{\cal A}_{1}+{\cal A}_{2}+{\cal A}_{3},\labell{44}\eeqa
where
\beqa
{\cal A}_{1}&\!\!\!\sim\!\!\!&2i\Tr(\lam_1\lam_2\lam_3)L_1\bigg\{\bigg[-\xi_{1i}\xi_{2j}k_{1a}k_{2b}
\Tr(P_{-}\fsH_{(n)}M_p\Gamma^{jbia})-p^ip^j\xi_{1i}\xi_{2j}\Tr(P_{-}\fsH_{(n)}M_p )\nonumber\\&&+\xi_{1i}\xi_{2j}p^ik_{2b}\Tr(P_{-}\fsH_{(n)}M_p \Gamma^{jb})+\xi_{1i}\xi_{2j}p^jk_{1a}\Tr(P_{-}\fsH_{(n)}M_p \Gamma^{ia})\bigg](-t-s-u-\frac{1}{2})
\nonumber\\&&
+\xi_1.\xi_2\Tr(P_{-}\fsH_{(n)}M_p)\bigg(
\frac{1}{2}(u+\frac{1}{4})(s+\frac{1}{4})\bigg)\bigg\},
\nonumber\\
{\cal A}_{2}&\sim&-2i\Tr(\lam_1\lam_2\lam_3)L_2 \bigg\{2k_{1a}k_{2b}\xi_1.\xi_2\Tr(P_{-}\fsH_{(n)}M_p \Gamma^{ba})\bigg\},
\nonumber\\
{\cal A}_{3}&\sim&-2i\Tr(\lam_1\lam_2\lam_3)L_2 \bigg\{-t\Tr(P_{-}\fsH_{(n)}M_p\Gamma^{ji})\xi_{1i}\xi_{2j} \bigg\}.
\eeqa

 The functions
 $L_1,L_2$ are :
\beqa
L_1&=&(2)^{-2(t+s+u)}\pi{\frac{\Gamma(-u+\frac{1}{4})
\Gamma(-s+\frac{1}{4})\Gamma(-t+\frac{1}{2})\Gamma(-t-s-u-\frac{1}{2})}
{\Gamma(-u-t+\frac{3}{4})\Gamma(-t-s+\frac{3}{4})\Gamma(-s-u+\frac{1}{2})}},\nonumber\\
L_2&=&(2)^{-2(t+s+u)-1}\pi{\frac{\Gamma(-u+\frac{3}{4})
\Gamma(-s+\frac{3}{4})\Gamma(-t)\Gamma(-t-s-u)}
{\Gamma(-u-t+\frac{3}{4})\Gamma(-t-s+\frac{3}{4})\Gamma(-s-u+\frac{1}{2})}}.
\nonumber\eeqa

 Note that all terms including transverse momentum of closed string $p^i,p^j$ are absent in the amplitude of one RR, two gauge field and one tachyon vertex operators \cite{Garousi:2007fk}. For their interpretations see section 4 of \cite{Hatefi:2012rx}.
 Now in order to compare the amplitude of \reef{44} with field theory, we are going to expand \reef{44} to actually produce all infinite number of tachyon and gauge poles . We postpone the expansion to the next section.
\\
\vskip 0.1in

Because of the presence  $\fsH_{(n)} $
and $\Gamma^{ba}$,  one believes  that the amplitude is non zero for $p=n-1$,  $p=n-3$ and $p=n+1$ cases. Taking a look at the poles in Gamma functions, one will observe that the  amplitude for $p=n+1$ has an infinite number of massless poles and so many contact interactions .
\\
$\frac{}{}$
\\
For the case of $p=n-1$, the amplitude has an infinite number of tachyon poles/contact terms. Whereas, for $p=n-3$ case, there are neither tachyon nor massless poles so the amplitude  does include just an infinite number of  contact terms. Now one may ask  how to expand this amplitude such that the leading terms of the amplitude correspond to the effective actions  and  non leading terms in our amplitude do belong to the higher derivative terms in the effective actions.

  In the next section, concerning momentum conservation along the world volume of brane, sending one Mandelstam variable $(t)$ to zero and sending the other two ones to mass square of tachyon, we conjecture  a unique expansion for all four point functions including one RR, massless fields and tachyon.
In particular, we will see that massless pole shows  that the kinetic term of the scalar field has no higher derivative correction as it has already been set in DBI action.

\section{Remarks on momentum expansion}

The momentum expansion of the amplitude must be achieved  by working out in detail either the tachyon or massless pole of field theory.
 Tachyon action has  $U(N)$ gauge
symmetry so one might search  a unique expansion for all four point functions including one RR, massless fields and tachyon. To discover this expansion,
we must remember the fact that  transverse scalar field and tachyon of
D$_8$-brane transform in the adjoint representation of $U(N)$
group. Therefore one may conclude that both of them should have the same non-abelian kinetic term, as it is the case.
Note that their  Feynman diagrams (taking their kinetic term into account) are also equal.
\\
$\frac{}{}$
\\
Recalling  the relation between Mandelstam variables and the momentum of RR field \reef{momencon},
 one does believe this on-shell constraint does not permit us to send all $s,t,u$ to zero.
Once more note that all on-shell conditions imply that the RR field must be non-zero, \ie $p_ap^a\rightarrow 1/4$. Thus we come to the point that computations make sense just for the non-BPS euclidean SD-brane.
Hence, once more  due to some kinematic reason \cite{Garousi:2008ge,Billo:1999tv} the S-matrix method must  be used to confirm all Wess-Zumino couplings only in the presence of  non-BPS SD-branes \cite{Gutperle:2002ai}. It is worth to talking about on-shell condition for RR 's momentum as follows:
\beqa
p_ip^i+p_ap^a=0
\nonumber\eeqa



For the definition of  SD$_p$-brane see \cite{Garousi:2008tn}.
Having used the kinematic relation \reef{momencon}, in which the
Mandelstam variables are satisfied, one can precisely reproduce all leading terms of the expansion with making use of the tachyon action.
Concerning  the fact that these relations  involve the mass of tachyon, one
really believes that the tachyon potential  should include  tachyon mass as well.

  There are two different assumptions in order to get the correct momentum expansion of a S-matrix element which are either   $(k_i+k_j)^2\rightarrow 0$ or $k_i\inn k_j\rightarrow 0$.  The case $(k_i+k_j)^2\rightarrow 0$ so happens  once we encounter a simple massless pole in $(k_i+k_j)^2$-channel. Applying this fact, we understand that our amplitude $(C\phi\phi T)$  just has a  simple massless pole in $(k_1+k_2)^2$-channel. Thus the correct momentum expansion in accord with the above argument must be done around
 \beqa
k_3.k_1\rightarrow 0,\qquad k_3.k_2\rightarrow 0,\qquad (k_1+k_2)^2\rightarrow 0,\label{esi12}\eeqa
Now we may use on-shell relations $k_1^2=k_2^2=0$ and $k_3^2=1/4$, to rewrite \reef{esi12}  as
\beqa
s\rightarrow -1/4,\qquad u\rightarrow -1/4,\qquad  t\rightarrow 0, \labell{point}\eeqa
 Therefore as it is clear, because of tachyon pole, the expansion is not usual $\alpha'$ expansion any more. So tachyon and massive modes will be decoupled.
 Our main goal is to find out new couplings for which precisely reproduce all terms of the expansion.

\vskip 0.1in
Regarding above remarks, the expansion of the function $L_1$ around  \reef{point} is

\beqa
L_1&=&-{\pi^{5/2}}\bigg( \sum_{n=0}^{\infty}c_n(s'+t+u')^n+\frac{\sum_{n,m=0}^{\infty}c_{n,m}[(s')^n(u')^m +(s')^m(u')^n]}{(t+s'+u')}\nonumber\\&&
+\sum_{p,n,m=0}^{\infty}f_{p,n,m}(s'+t+u')^p[(s'+u')^n(s'u')^{m}]\bigg)
\labell{high6587},\eeqa
Note that  the structure of  contact interaction terms  in \reef{high6587} is not the same. Hence, we believe contact terms in the second line of  \reef{high6587} are related to different couplings as we will derive them in the field theory side as well. They were corresponding to different field theory couplings. The expansion of the function $L_2$ around \reef{point} is
\beqa
L_2&=&-\pi^{3/2}\bigg(\frac{1}{t}\sum_{n=-1}^{\infty}b_n(u'+s')^{n+1}+\sum_{p,n,m=0}^{\infty}e_{p,n,m}t^{p}(s'u')^{n}(s'+u')^m\bigg).\labell{high67}\eeqa
where in the above relations we have considered $u'=u+\frac{1}{4}=-\alpha'k_2.k_3$ and $s'=s+\frac{1}{4}=-\alpha'k_1.k_3$. One important point that should be made is that  all $b_n$ coefficients  are the same as those appeared  in the momentum expansion of the amplitude  of one RR, three massless open string  vertex operators for BPS branes \cite{Hatefi:2010ik,Hatefi:2012ve,Hatefi:2012rx}. Let us write some of them down
\beqa
&&b_{-1}=1,\,b_0=0,\,b_1=\frac{1}{6}\pi^2,\,b_2=2\z(3),\nonumber\\
&&e_{2,0,0}=e_{0,1,0}=2\z(3),e_{1,0,0}=\frac{1}{6}\pi^2,e_{1,0,2}=\frac{19}{60}\pi^4,e_{1,0,1}=e_{0,0,2}=6\z(3),\nonumber\\
&&e_{0,0,1}=\frac{1}{3}\pi^2,e_{3,0,0}=\frac{19}{360}\pi^4,e_{0,0,3}=e_{2,0,1}=\frac{19}{90}\pi^4,e_{1,1,0}=e_{0,1,1}=\frac{1}{30}\pi^4,\labell{577}\\
&&c_0=0,c_1=\frac{\pi^2}{3},c_2=4\z(3),
\,c_{1,1}=\frac{\pi^2}{3},\,c_{0,0}=1,c_{3,1}=c_{1,3}=\frac{4\pi^4}{15},c_{2,2}=\frac{2\pi^4}{15},\nonumber\\
&&c_{1,0}=c_{0,1}=0
,c_{3,0}=c_{0,3}=0\,
\,c_{2,0}=c_{0,2}=\frac{\pi^2}{3},c_{1,2}=c_{2,1}=-8\z(3),\nonumber\\
&&f_{0,1,0}=-\frac{2\pi^2}{3},\,f_{0,2,0}=-f_{1,1,0}=12\z(3),f_{0,0,1}=4\z(3)\,c_{4,0}=c_{0,4}=\frac{2\pi^4}{15}, \nonumber
\eeqa

\vskip 0.05in

However, some of the coefficients, in particular $c_{n,m}$ are different from those that appeared in \cite{Hatefi:2010ik} because
the expansions are not the same. For the amplitude of $CAAA$ the expansion was low energy expansion while here the expansion is not low energy expansion  even though apparently they do have the same structure. We will see that these coefficients play a key role in confirming the infinite number of tachyon poles in string theory. Now it becomes clear that $L_1$ has an infinite number of tachyon poles in $s'+t+u'$-channel and $L_2$ has an infinite number of  massless poles in $t$-channel. These poles must  be reproduced in field theory by well defined couplings. In the next section, we talk about effective field theory, then we try to produce the first simple tachyon pole within field theory context, then we go on to find all of the infinite numbers of the couplings between two scalar fields and two tachyons to all orders of $\alpha'$, which can be approved just by S-matrix computations.

\section{ Effective Field Theory on the World-Volume}

On-shell states  in our amplitude are   in fact
 open string  tachyon and  scalar fields which are coming from DBI action  and the closed string RR field
which is coming from the Wess-Zumino action. One may wonder whether there could be an off-shell gauge field in our amplitude. We confirm its existence in the field theory side. We are also working  in flat
background. Applying square root expansion mentioned in \cite{Hatefi:2010ik}
we could find  non abelian kinetic terms as

 \beqa {\cal
L}&=&-T_p(\pi\alpha')\Tr\left(m^2T^2+D_aTD^aT-(\pi\alpha')F_{ab}F^{ba}\right),\labell{expandL2}
\eeqa
In order to actually have  gauge field and tachyon propagators we need to keep them as well.


On the other hand, the  scalar fields have geometrical meaning which represent transverse coordinates of
the brane in  effective theory.

One has to mention that scalars have $U(N)$
gauge symmetry and they  have to be in the adjoint  representation. In general scalars do appear in the effective action in three different ways  and we now list them .
\vskip 0.1in

In the first way, they have been confirmed in Wess-Zumino action in its exponential  \reef{finalcs}. So our notation is as follows

\beqa
i_\Phi i_\Phi C^{(n)}={1\over2(n-2)!}[\Phi^i,\Phi^j]\,
C^{(n)}_{ji\mu_3\cdots\mu_n}dx^{\mu_3}\cdots dx^{\mu_n},\quad C^{(n)}={1\over n!}C^{(n)}_{\mu_1\cdots\mu_n} dx^{\mu_1}\cdots dx^{\mu_n}
 \labell{inside}\eeqa

In the second way, we will have the
covariant derivatives of the non abelian scalar fields in  pullback as follows
\beqa
P[E]_{ab}=E_{ab}+\lambda\,E_{ai}\,D_{b}\Phi^i+
\lambda\,E_{ib}\, D_a\Phi^i+
\lambda^2\,E_{ij}D_a\Phi^iD_b\Phi^j,
\eeqa

In the last way,the metric will be given by a non-abelian Taylor expansion as
 \beqa
G_{\mu\nu}&=&\exp\left[\lambda\Phi^i\,{\prt_{x^i}}\right]G^0_{\mu\nu}
(\sigma^a,x^i)|_{x^i=0}
\labell{slick}\nonumber\\
&=&\sum_{n=0}^\infty {\lambda^n\over n!}\,\Phi^{i_1}\cdots\Phi^{i_n}\,
(\prt_{x^{i_1}}\cdots\prt_{x^{i_n}})G^0_{\mu\nu}
(\sigma^a,x^i)|_{x^i=0}\ .
\eeqa

Hence, one might reveal that the action does involve all transverse
 derivatives of the closed string  RR fields within Taylor expansion.

$\frac{}{}$
\\
The existence of the  scalar fields in  pullback \cite{Hull:1997jc} and the functional
dependence \cite{Douglas:1997zw} have already been addressed.
Here we would like to establish the explicit existence of the commutator interactions of scalar fields even to tachyon in
Wess-Zumino action. Another ambitious goal is to  obtain some sort of new couplings which will be investigated
by honest string scattering computations  which
 will be explained in the section of contact terms.
 In order to actually have some non trivial couplings, we have to have at least
 three kinds of open string states and a single closed string state. One may point out that the leading terms for  scalar fields could have been found  from Born-Infeld action as
\beq
-{\lambda^2T_p}\int d^{p+1}\sigma\,\Tr\left({1\over2}D^a\Phi^iD_a\Phi^i
-{1\over4}[\Phi^i,\Phi^j][\Phi^i,\Phi^j]\right)\,,
\labell{scalactt}
\eeq
One more thing which is really worth saying is that pullback of closed string fields must be defined in the static
gauge. Also notice that in the pullback we must have covariant derivatives of non abelian scalars as
\
\beqa
P[\eta_{ab}]=\eta_{ab}+2\pi\alpha'D_a\phi^iD_b\phi^j\eta_{ij}.
\nonumber\eeqa
The first term in \reef{scalactt} is indeed non abelian kinetic term of scalar fields which can be regarded as a reduction of
$F^2$ in
ten dimensions.
\vskip 0.1in

Within our conventions, $D_a\Phi^i=\prt_a\Phi^i+i[A_a,\Phi^i]$.
Note that the non abelian scalar field theory now does involve the interactions with
even gauge fields, which are either inside of
 three-point function $\Tr(\prt^a\Phi^i\,[A_a,\Phi^i])$ or four-point function $\Tr([A^a,\Phi_i][A_a,\Phi^i])$.
So we expect from the field theory point of view, Feynman diagrams in which two  open
scalars  may scatter to give an off-shell gauge field
 which is going to be  absorbed by a lower order RR coupling (in the bulk) and one external tachyon in the world volume space. This does happen for $p=n+1$ case. Thus  we will see in field theory this diagram is
responsible for an infinite number of massless gauge poles indicating in the string amplitude. We also determine all their contact terms .

Notice the fact that due to the appearance of unusual kinematics and the mass of tachyon we are not able to derive our amplitude from an usual five-point open string amplitude. That is why we have done  the computations of this amplitude  directly, which are indeed more sophisticated than five-point pure open string amplitude .

\subsection{$p=n-1$ case}
Having set the usual
Chern-Simons action, it seems that there is no coupling between RR fields of the type II string theory
and the non-BPS D-branes. However, one must point out that there is a nonvanishing coupling between
the RR field and one tachyon on the world-volume of these branes,
such that the
Chern-Simons action was modified in \cite{Billo:1999tv} to be responsible for this coupling.

In this section we would like to obtain the first tachyonic pole of the amplitude for $p=n-1$ case  and then we proceed to discover an infinite number of higher derivative corrections that are related to two tachyons and two scalar fields. In order to show that our proposal for  these infinite numbers of couplings works  we use these couplings to check  the infinite number of tachyon poles of the amplitude for  $p=n-1$ case later on. Finally we will consider all contact terms together. Only the last term in ${\cal A}_1$ in \reef{44} is related to singular terms. Therefore  for this case the trace is:
\beqa
\Tr\bigg(\fsH_{(n)}M_p
\bigg)&=&\pm\frac{32}{n!}\eps^{a_{0}\cdots a_{p}}H_{a_{0}\cdots a_{p}}
,\nonumber\eeqa
 Note that the trace also does involve the factor of $\ga^{11}$ and keeps held  the following
results for $p>3$ with $H_{(n)} \equiv \ast H_{(10-n)}$ for
$n\geq 5$.
Replacing  this trace in the last term of  ${\cal A}_1$,  one gets
\beqa
{\cal A}^{\phi\phi TC}&=&\frac{\mp 32i}{(p+1)!}(\beta'\mu'_p\pi^{1/2})\Tr(\lambda_1\lambda_2\lambda_3)\eps^{a_{0}\cdots a_{p}}H_{a_{0}\cdots a_{p}}L_1\bigg\{
\frac{1}{2}(\xi_1.\xi_2)(u+\frac{1}{4})(s+\frac{1}{4})\bigg\}
\labell{1pptc}
\eeqa
where  $(\beta'\mu'_p\pi^{1/2})$ is a normalization factor. As it is clear  \reef{1pptc} is symmetric under changing massless scalar fields, therefore the amplitude is non-zero even for the abelian gauge group.
 Notice that we do not want to fix  the sign of the amplitudes.

It is seen that \reef{1pptc} has an infinite number of tachyon poles in the $(s'+t+u')$-channel and does include an infinite number of contact terms. In the next section we would like to show that just the first tachyon pole in \reef{1pptc} will be reproduced by applying the symmetric trace prescription of  tachyonic DBI action. However, in order to be able to produce an infinite number of tachyon poles of the desired amplitude up to all orders of $\alpha'$, one must find an infinite number of higher derivative corrections of two scalar fields and two tachyons. Making use of their explicit forms and extracting the vertex of two on-shell scalars, one tachyon and one off-shell  tachyon  in the world-volume theory  of N coincident  non-BPS branes, we will be able to produce all infinite number of tachyon poles in \reef{1pptc} in a precise manner.

\subsection{First Tachyon pole for $p=n-1$ case}

 Here we are going to mention the general form of tachyonic action \cite{Garousi:2008nj,Garousi:2008ge} as the following:

 \beqa
S_{DBI}&=&-\frac{T_p}{2}\int
d^{p+1}\sigma \STr\left(V({ T_iT_i})\sqrt{1+\frac{1}{2}[T_i,T_j][T_j,T_i]}\right.\nonumber \\
&&\qquad\qquad\left.
\times\sqrt{-\det(\eta_{ab}
+2\pi\alpha'F_{ab}
+2\pi\alpha'D_a{ T_i}(Q_{1}^{-1})_{ij}D_b{ T_j})} \right)\,,\label{final 1action}\eeqa

where the tachyon potential appeared in \reef{tpot} and
\beqa
Q_{1ij}&=&I\delta_{ij}-i[T_i,T_j],\eeqa
We must take into account the fact that the trace in the action
is totally symmetric between
 $F_{ab},D_aT^i$,
 $[T^i,T^j]$ and  $T^i$ inside the potential $V(T^iT^i)$.

  One can write down  $V(T)$  around its
maximum which is $T_{max}=0$ so that \beqa
V(T)=1-\frac{\pi}{2}T^2+\frac{\pi^2}{8}T^4+O(T^6)\,\,.\labell{finalv}
\eeqa
As observed with  A.Sen 's
conjecture the coefficient of $T^4$ is a consistent result which indicates  that the tachyon potential has a minimum
 \cite{Sen:1998ii,Sen:1999mh}. Therefore one can see that by neglecting  $O(T^6)$ terms,
 $V(T)$ does have a minimum at $T_{min}=\sqrt{2/\pi}$, where the value of the potential at its minimum is $V(T_{min})=0.5$.

 So we conclude that $V(T_{min})$ is now non zero; however, at real stable vacuum the value of the potential must be zero
 so $V(T_{min})=0.5$ means that one must consider  the higher order
$O(T^6)$ terms in the potential. The other point we want to make  is that, A.Sen
showed that  the minimum of the  potential is at $T_{min}\rightarrow \infty$, and $V(T)$ has exponential behaviour just like $e^{-\sqrt{\pi}T}$
 around its minimum  \cite{Sen:2002an}.

\vskip 0.1in

 However within our formalism we want to study this action  around the unstable point $T=0$ and around $T\rightarrow\infty$ just the second term in \reef{finalv} is dominant. Thus, above action
does reduce to usual action with potential $T^4V(T^2)$ and at $T\rightarrow\infty$ this potential tends to zero. This is already expected
from tachyon condensation for a single non-BPS brane.
 To check the consistency \reef{final 1action} with some disk level amplitudes in super string theory see \cite{Garousi:2007fk,Garousi:2008nj} .

\vskip 0.1in

Applying the expansion of the square
root of determinant $\sqrt{det(M_0+M)}$ \cite{Hatefi:2010ik} and making use of both actions \reef{finalbi} and \reef{final 1action}, one gets various interactions.
 However, we are interested in considering
 two scalar and two tachyon couplings thus we will have  the following couplings:

 \beqa
{\cal{L}}(\phi,\phi,T,T)&=& -2T_p(\pi\alpha')^3{\rm
STr} \bigg(
m^2\cT^2(D_a\phi^iD^a\phi_i)+\frac{\alpha'}{2}D^{\alpha}\cT D_{\alpha}\cT D_a\phi^iD^a\phi_i \nonumber\\&&-
\alpha' D^{b}\cT D^{a}\cT D_a\phi^iD_b\phi_i \bigg)\labell{dbicoupling}. \eeqa

 Note that in the above couplings one tachyon has to be off-shell and the external states are two scalars and one tachyon in which  they must carry their own momentum. That is why the coupling $\frac{1}{4\pi\alpha'}[X^i,T][X_i,T]$ has been over looked. One more point  to mention is that  after averaging all
possible permutations in \reef{dbicoupling}  one has to take also an overall trace on the group theory.
We want to obtain the higher derivative couplings of the two scalars and two tachyons
 and then investigate that
 these terms reproduce an infinite number of tachyon poles in the amplitude,
 and finally we end up with some new couplings which can be confirmed just by direct scattering amplitude
 computations.

Because of the fact that propagator is Abelian,  we need to consider two possible orderings in order to get $\Tr(\lambda_1\lambda_2\lambda_3)$ ordering. Having written down symmetric traces in terms of ordinary traces, one can essentially find out  the terms   $(L_1^{0,0}+L_2^{0,0}+L_3^{0,0}+L_4^{0,0})$  such that
\beqa
\cL_{}&=&-2T_p(\pi\alpha')^3(\cL_{1}^{0,0}+\cL_{2}^{0,0}+\cL_{3}^{0,0}+\cL_{4}^{0,0}),\labell{3lagrang}\eeqa
where
\beqa
\cL_1^{0,0}&=&\frac{-4m^2}{\pi^2}
\Tr\bigg(\frac{}{}a_{0,0}(\cT^2 D_a\phi^iD^a\phi_i)+\frac{}{}b_{0,0}(\cT D_a\phi^i\cT D^a\phi_i)\bigg),\nonumber\\
\cL_2^{nm}&=&\frac{-4}{\pi^2}\Tr\left( \frac{}{} a_{0,0}(D^{\alpha}\cT D_{\alpha}\cT D_a\phi^iD^a\phi_i)+
\frac{}{}b_{0,0}(D^{\alpha} \cT D_a\phi^i D_{\alpha}\cT D^a\phi_i)\right),\nonumber\\
\cL_3^{nm}&=&\frac{4}{\pi^2}\Tr
\left(\frac{}{}a_{0,0}(D^{\beta}\cT D_{\mu}\cT D^\mu\phi^iD_\beta\phi_i)+\frac{}{}b_{0,0}(D^{\beta}\cT D^\mu\phi^iD_{\mu}\cT D_\beta\phi_i)\right),\nonumber\\
\cL_4^{nm}&=&\frac{4}{\pi^2}\Tr\left(\frac{}{}a_{0,0}(D^{\beta}\cT D^{\mu}\cT D_\beta\phi^iD_\mu\phi_i)
+\frac{}{}b_{0,0}(D^{\beta}\cT D_\beta\phi^iD^{\mu}\cT D_\mu\phi_i)\right).\labell{higher0}
\eeqa
with $a_{0,0}=\frac{-\pi^2}{6},b_{0,0}=\frac{-\pi^2}{12}$.
Let us mention the useful relation
$\STr(T^2LL)=\frac{2}{3}\Tr(TTLL)+\frac{1}{3}\Tr(TLTL)$.
Thus the first tachyon  pole of the amplitude \reef{1pptc} may be found as
\beqa
{\cal A}&=&V_{\alpha}(C_{p},T)G_{\alpha\beta}(T)V_{\beta}(T,T_3,
\phi_1,\phi_2),\labell{ampu549}\eeqa
The tachyon propagator and the vertex $V_{\alpha}(C_{p},T)$ which have no higher derivative correction might be appeared as

\beqa
G^{\alpha\beta}(T) &=&\frac{i\delta^{\alpha\beta}}{(2\pi\alpha') T_p
(-k^2-m^2)},\nonumber\\
V^{\alpha}(C_{p},T)&=&2i\mu'_p\beta'(2\pi\alpha')\frac{1}{(p+1)!}\epsilon^{a_0\cdots a_{p}}H_{a_0\cdots a_{p}}\Tr(\Lambda^{\alpha}).
\labell{Fey55}
\eeqa
  $\alpha,\beta$ are group indices. Note that in \reef{Fey55}, $\Tr(\Lambda^{\alpha})$ is nonzero just for Abelian matrix $\Lambda^{\alpha}$.
The vertex $ V^{\beta}(T,T_3,\phi_1,\phi_2)$  can be  derived from \reef{3lagrang}. Recalling that  off-shell tachyon must be Abelian, one gets
\beqa
&&V^{\beta}(T,T_3,\phi_1,\phi_2)=4iT_p(\pi^3\alpha')(\alpha')^{2}\Tr(\lambda_1\lambda_2\lambda_3\Lambda^{\beta})\bigg[\frac{1}{2}(\xi_1.\xi_2)
(u+\frac{1}{4})(s+\frac{1}{4})\bigg],\labell{veraattm}\eeqa
  Note that in order to obtain this vertex we have to consider all two group factors  of $\Tr(\lambda_3\Lambda^{\beta}\lambda_1\lambda_2)$ and  $\Tr(\Lambda^{\beta}\lambda_3\lambda_1\lambda_2)$.

However, the other terms  will have the contribution of  $\Tr(\lambda_2\lambda_1\lambda_3\Lambda^{\beta})$  to \reef{veraattm}.
 Having replaced   \reef{veraattm} in the Feynman amplitude \reef{ampu549}, we get the following result in the field theory side:
\beqa
&&-64i\pi^3\beta'\mu_p'\frac{ \eps^{a_{0}\cdots a_{p}}H_{a_{0}\cdots a_{p}}}{(p+1)!(s'+t+u')}\Tr(\lam_1\lam_2\lam_3)
\bigg[(\xi_1.\xi_2)\frac{1}{2}u's'
\bigg].\labell{amphigh7}\eeqa
Notice that we just considered the first tachyon pole of $L_1$, which had the coefficient of $2\pi^\frac{5}{2}$.  Thus symmetrized trace prescription of tachyonic DBI action could precisely reproduce the first tachyon pole of string theory amplitude. How can we produce the other tachyon poles? In detail below we obtain all  couplings between two tachyons and two scalars to all orders in $\alpha'$ and then use them we produce all infinite numbers of tachyon poles for $p=n-1$ case in the field theory side as well.
\section{Two scalar and two tachyon couplings up to all orders of $\alpha'$}

It is known the higher derivative corrections  do have some kind of field redefinition freedom.
 Therefore we may be able to pick this freedom up and relate it
to  some couplings in field theory. For example  we can relate $(\alpha')^3$ terms
  to the couplings that involve $\prt\prt\prt T$ and so on.

 It is indeed an interesting issue to find out  these higher derivative terms for
 all orders of $\alpha'$ which is one of the main goals of the paper.
 \\

A precise method for obtaining the general form of higher derivative theories is indeed studying in detail the S-matrix method.
Using this method one first has to find S-matrix elements of the desired theory
 and then try to compare them with the S-matrix elements of super string theory. If these higher derivative  terms are equal with the string theory, then their S-matrix elements   must be the same with the momentum expansion of the  S-matrix elements of string theory. Therefore,
 in order to get those couplings, we should look for S-matrix elements and try to expand them in such a way that the correct  higher derivative couplings are discovered in field theory.  As an example, in order to get the S-matrix elements of two real tachyons and  one closed string RR field, the following higher derivative couplings in brane-antibrane   must be taken into account:
\beqa
2i\alpha'\mu_p\sum_{n=0}^{\infty}a_n\left(\frac{\alpha'}{2}\right)^n C_{p-1}\wedge (D ^aD_a)^n(DT\wedge DT^*).\labell{hderv}
\eeqa
For more information see \cite{Garousi:2007fk,Garousi:2008xp}.

However, for four and five-point functions, it is really a complicated task to find higher derivative corrections. Let us address an issue. The S-matrix elements
of   tachyons  might not have definite  physical meaning; nevertheless,
if the two theories are equal then we believe that the string theory S-matrix elements  must be reproduced just by their higher derivative  couplings.

One important point on tachyonic action is that
it produces just leading terms of
S-matrix elements at the top of  tachyon potential
(not $\alpha'\rightarrow 0$ limit). Therefore we may expect already that the
other terms of expansion really  have important effect.  So one concludes that in field
theory effective action must have all those  higher derivative terms as well.
Let us come to our main point.
\vskip 0.1in

  In this section we would like to obtain the  infinite number of the couplings between  two scalars and two tachyons  in order to reproduce  the string theory S-matrix elements to all orders of $\alpha'$.


Here we just mention our method for finding the higher derivative extensions of the above couplings \reef{higher0}. Two important points are worth mentioning. The first point is that the kinematic factor in the amplitude of two tachyons and two scalar fields, is
 the vertex  of two on-shell tachyons and two on-shell scalar fields of \reef{higher0} and the second point is that,
 the coefficient of all higher order terms in the amplitude of two tachyons and two scalar fields is exactly the vertex of two tachyons and two scalars. Thus we might find out  the higher derivative couplings by acting  suitable higher  derivatives  on these couplings \reef{higher0}.  Each term in the above couplings has the coefficient  $a_{0,0}$ and $b_{0,0}$. In the higher derivative orders  one must substitute them by  $a_{n,m}$ and $b_{n,m}$. For further details see \cite{Hatefi:2010ik,Hatefi:2012ve,Hatefi:2012rx,Garousi:2008ge,Garousi:2008xp}.

$\frac{}{}$

Concerning T-duality transformation, one may expect that the higher derivative couplings of two scalar fields and two tachyons might be similar to the higher derivative couplings of two gauge fields and two tachyons.
To realize the differences for gauge and scalar field see section 5 of \cite{Hatefi:2010ik} .

Because of two important facts one may try to apply the general form of T-duality transformation to discover  higher derivative  couplings  two scalar fields and two  tachyons on the world
volume of $N$ non-BPS D-branes, to all orders of $\alpha'$ and then precisely fix their coefficients comparing with S-matrix elements.

\vskip 0.1in

Let us make some comments.
 The first comment is that the Mandelstam variables for the amplitudes of $TT\phi\phi$  and $TTAA$ must satisfy  the same  constraint. Notice that all massless poles coming from the non-Abelian kinetic term have to be reproduced by sending some of the Mandelstam variables to zero. The second point that should be made is that, all external states  in those two amplitudes satisfy the same on-shell and physical state condition .

The method for finding those infinite numbers of couplings between two gauge fields and two tachyons has been mentioned in \cite{Garousi:2008ge,Garousi:2008xp}.

\vskip 0.2in


The massless poles in the amplitude of $TTAA$ have been reproduced by the non-Abelian kinetic term  of the gauge fields and their Lagrangian is

\beqa
{\cal{L}}(A,A,T,T)&=&(\pi\alpha')^3{\rm
STr} \left(
\frac{}{}m^2\cT^2F_{\mu\nu}F^{\nu\mu}+\frac{}{}D^{\alpha}\cT D_{\alpha}\cT F_{\mu\nu}F^{\nu\mu}-
4F^{\mu\alpha}F_{\alpha\beta}D^{\beta}\cT D_{\mu}\cT
\right).\nonumber
\eeqa
However, for  the amplitude of $TT\phi\phi$ the massless poles are reproduced by the non-Abelian kinetic term  of the scalar fields and their Lagrangian  appears in \reef{dbicoupling}.

If we compare the above Lagrangian with \reef{dbicoupling} then we realize that they have some differences in the  the  indices and coefficients.
In order to replace $F$ with $D\phi$ and to observe more details see section 5 of \cite{Hatefi:2010ik}.

To avoid further details we just write down the results that we found using direct S-Matrix computations.

 Therefore by extracting symmetrized traces in terms of usual traces we were able to find out the couplings between two tachyons and two covariant derivative of scalar fields on the world volume of $N$ coincident non-BPS D-branes, to all orders of $\alpha'$  as the following:

\beqa
\cL_{}&=&-2T_p(\pi\alpha')(\alpha')^{2+n+m}\sum_{n,m=0}^{\infty}(\cL_{1}^{nm}+\cL_{2}^{nm}+\cL_{3}^{nm}+\cL^{nm}_{4}),\labell{lagrango}\eeqa
where
\beqa
\cL_1^{nm}&=&m^2
\Tr\left(\frac{}{}a_{n,m}[\cD_{nm}(\cT^2 D_a\phi^iD^a\phi_i)+ \cD_{nm}(D_a\phi^iD^a\phi_i\cT^2)]\right.\nonumber\\
&&\left.+\frac{}{}b_{n,m}[\cD'_{nm}(\cT D_a\phi^i\cT D^a\phi_i)+\cD'_{nm}( D_a\phi^i\cT D^a\phi_i\cT)]+h.c.\right),\nonumber\\
\cL_2^{nm}&=&\Tr\left(\frac{}{}a_{n,m}[\cD_{nm}(D^{\alpha}\cT D_{\alpha}\cT D_a\phi^iD^a\phi_i)+\cD_{nm}( D_a\phi^iD^a\phi_i D^{\alpha}\cT D_{\alpha}\cT)]\right.\nonumber\\
&&\left.+\frac{}{}b_{n,m}[\cD'_{nm}(D^{\alpha} \cT D_a\phi^i D_{\alpha}\cT D^a\phi_i)+\cD'_{nm}( D_a\phi^i D_{\alpha}\cT D^a\phi_i D^{\alpha} \cT)]+h.c.\right),\nonumber\\
\cL_3^{nm}&=&-\Tr
\left(\frac{}{}a_{n,m}[\cD_{nm}(D^{\beta}\cT D_{\mu}\cT D^\mu\phi^iD_\beta\phi_i)+\cD_{nm}( D^\mu\phi^iD_\beta\phi_iD^{\beta}\cT D_{\mu}\cT)]\right.\nonumber\\
&&\left.+\frac{}{}b_{n,m}[\cD'_{nm}(D^{\beta}\cT D^\mu\phi^iD_{\mu}\cT D_\beta\phi_i)+\cD'_{nm}(D^\mu\phi^i D_{\mu}\cT  D_\beta\phi_i  D^{\beta}\cT)]+h.c.\right),\nonumber\\
\cL_4^{nm}&=&-\Tr\left(\frac{}{}a_{n,m}[\cD_{nm}(D^{\beta}\cT D^{\mu}\cT D_\beta\phi^iD_\mu\phi_i)
+\cD_{nm}( D^\beta\phi^iD^\mu\phi_iD_{\beta}\cT D_{\mu}\cT)]\right.\nonumber\\
&&\left.+\frac{}{}b_{n,m}[\cD'_{nm}(D^{\beta}\cT D_\beta\phi^iD^{\mu}\cT D_\mu\phi_i)+\cD'_{nm}( D_\beta\phi^i D_{\mu}\cT  D^\mu\phi_i D^{\beta}\cT)]+h.c.
\right),\nonumber
\eeqa
   If we did calculate  the coupling of two on-shell tachyons and two on-shell scalar fields from \reef{lagrango}, one would be able to
   talk about all the interactions  in the amplitude of two scalar fields and two tachyons.
   The higher derivative operator $\cD_{nm}$ and $\cD'_{nm}$ may be read as
\beqa
\cD_{nm}(EFGH)&\equiv&D_{b_1}\cdots D_{b_m}D_{a_1}\cdots D_{a_n}E  F D^{a_1}\cdots D^{a_n}GD^{b_1}\cdots D^{b_m}H, \nonumber\\
\cD'_{nm}(EFGH)&\equiv&D_{b_1}\cdots D_{b_m}D_{a_1}\cdots D_{a_n}E   D^{a_1}\cdots D^{a_n}F G D^{b_1}\cdots D^{b_m}H.\eeqa
 So \reef{lagrango} is the higher derivative correction of two scalar field and two tachyon couplings of tachyonic action. One important evidence in confirming \reef{lagrango} is that by setting  the covariant derivative of the scalar field and the second covariant derivative of tachyon to zero, \reef{lagrango}  goes back to the couplings \reef{dbicoupling}.
 This definitely shows  that  when fields vary  so slowly, the non-Abelian tachyon DBI action is going to be the correct  effective action for  non-BPS SD-branes.
\\

One extremely important fact about the couplings in \reef{lagrango} is that  they may  have on-shell ambiguity, which means that there is no difference between $T$ and $ 2\alpha'D_aD^a T$ because they do have an identical effect. However, this ambiguity does not have any  effect on the simple massless and tachyon  poles  of the amplitude. In the case of massless poles it is sort of an obvious thing because the tachyons are on-shell. In the case of tachyon poles we will show that
 $k^2+m^2$ terms  cancelled  by tachyon pole, and eventually one gets some contact terms. Thus the difference is just  an extra contact interaction. By doing an amplitude where the couplings \reef{lagrango} would appear either in tachyon poles or contact terms, one will be able to fix that ambiguity in \reef{lagrango}. These couplings might appear in the tachyonic pole of S-matrix elements of two tachyons, two scalar fields and one gauge field. It would be nice to follow in detail this amplitude, in favor of those couplings. This amplitude does have a long computation
that we leave  for future work
\cite{Hatefi:2012eh}.
\\

In the next section we will show that an infinite number of the tachyon
 poles  will  result by taking the Wess-Zumino coupling $C_{p}\wedge DT$ and by our proposed  higher
 derivative couplings  of two scalar fields and two tachyons,  which we discovered  in \reef{lagrango}.

\subsection{Infinite number of tachyon poles for $p+1=n$ case }
As we promised in this section we are going to investigate that the infinite number of two scalar field and two tachyon couplings  in \reef{lagrango}  will result in the infinite number of tachyon poles of the string theory amplitude in the $(s'+t+u')$-channel.
 To do so, we must take the following Feynman diagram where one RR in the world volume of non-BPS  branes decays to one  tachyon and two scalar fields as follows :

\beqa
{\cal A}&=&V^{\alpha}(C_{p},T)G^{\alpha\beta}(T)V^{\beta}(T,T_3,\phi_1,\phi_2),\labell{amp544u}\eeqa
 In order to proceed one needs the vertex of one RR p-form field and one off-shell tachyon  and tachyon propagator as
\beqa
G^{\alpha\beta}(T) &=&\frac{i\delta^{\alpha\beta}}{(2\pi\alpha') T_p
(-k^2-m^2)},\nonumber\\
V^{\alpha}(C_{p},T)&=&2i\mu'_p\beta'(2\pi\alpha')\frac{1}{(p+1)!}\epsilon^{a_0\cdots a_{p}}H_{a_0\cdots a_{p}}\Tr(\Lambda^{\alpha}).
\labell{Fey}
\eeqa
Note that $\Tr(\Lambda^{\alpha})$ makes sense just for the Abelian gauge group. One also needs to find out the vertex of one off-shell,one on-shell tachyon and two on-shell scalar fields. This vertex must be obtained from (\ref{lagrango}). Regarding off-shell tachyon as the Abelian one, we get the vertex of $V^{\beta}(T,T_3,\phi_1,\phi_2)$  as follows

\beqa
&&2iT_p(\pi\alpha')(\alpha')^{2+n+m}
(a_{n,m}+b_{n,m})\Tr(\lambda_1\lambda_2\lambda_3\Lambda^{\beta})\bigg[\frac{1}{2}(\xi_1.\xi_2)(u+\frac{1}{4})(s+\frac{1}{4})\bigg]\nonumber\\&&\times\bigg((k_3\inn k_1)^n(k_3\inn k_2)^m+(k_3\inn k_1)^n(k_1\inn k)^m
+(k\inn k_2)^m(k\inn k_1)^n+(k_1\inn k)^n(k_3\inn k_1)^m\nonumber\\&&+(k_3\inn k_2)^m(k_2\inn k)^n+(k\inn k_2)^n(k_1\inn k)^m
+(k_3\inn k_2)^n(k_1\inn k_3)^m+(k_3\inn k_2)^n(k_2\inn k)^m\bigg).\labell{veraatt}\eeqa
In \reef{veraatt}  $k$ does indicate the off-shell tachyon's momentum. Note that in order to obtain this vertex we must consider both possible cases as the following :
\beqa
\Tr(\lambda_3\Lambda^{\beta}\lambda_1\lambda_2),\quad
\Tr(\Lambda^{\beta}\lambda_3\lambda_1\lambda_2).
\nonumber
\eeqa
Other cases have different coefficients, like  $\Tr(\lambda_2\lambda_1\lambda_3\Lambda^{\beta})$, and should  not be computed as they made no contribution  to the desired vertex. Because of the fact that we need  some of the coefficients $a_{n,m}$ and $b_{n,m}$ , let us address some of them as  \cite{Hatefi:2010ik,Garousi:2008xp}
\beqa
&&a_{0,0}=-\frac{\pi^2}{6},\,b_{0,0}=-\frac{\pi^2}{12},a_{1,0}=2\z(3),\,a_{0,1}=0,\,b_{0,1}=-\z(3),a_{1,1}=a_{0,2}=-7\pi^4/90,\nonumber\\
&&a_{2,2}=(-83\pi^6-7560\z(3)^2)/945,b_{2,2}=-(23\pi^6-15120\z(3)^2)/1890,a_{1,3}=-62\pi^6/945,\nonumber\\
&&\,a_{2,0}=-4\pi^4/90,\,b_{1,1}=-\pi^4/180,\,b_{0,2}=-\pi^4/45,a_{0,4}=-31\pi^6/945,a_{4,0}=-16\pi^6/945,\nonumber\\
&&a_{1,2}=a_{2,1}=8\z(5)+4\pi^2\z(3)/3,\,a_{0,3}=0,\,a_{3,0}=8\z(5),b_{1,3}=-(12\pi^6-7560\z(3)^2)/1890,\nonumber\\
&&a_{3,1}=(-52\pi^6-7560\z(3)^2)/945, b_{0,3}=-4\z(5),\,b_{1,2}=-8\z(5)+2\pi^2\z(3)/3,\nonumber\\
&&b_{0,4}=-16\pi^6/1890.\eeqa
where $b_{n,m}$ should be  symmetric.

The following relations must be pointed out.
 \beqa
 k_1\inn k=k_2.k_3+(-k^2-m^2)/2, \quad k_2\inn k=k_1.k_3+(-k^2-m^2)/2 \nonumber\eeqa

Note that $-k^2-m^2$ in the  vertex \reef{veraatt} will be removed with those common terms in the propagator and will give rise to some contact interactions of one RR, two scalar fields and one tachyon that we do not consider   now. So for the moment we do over look them. In fact, we will observe that finally one has to subtract them from  the interaction terms that concluded  from the amplitude of one RR, two scalar fields and one  tachyon. In the last section once again we try to come back to those terms. Having neglected them, one gets an infinite number of tachyon poles as  :
\beqa
&&-32i\pi\alpha'^{2}\beta'\mu_p'\frac{ \eps^{i_{0}\cdots i_{p}}H_{i_{0}\cdots i_{p}}}{(p+1)!(s'+t+u')}\Tr(\lam_1\lam_2\lam_3)
\sum_{n,m=0}^{\infty}\bigg((a_{n,m}+b_{n,m})[s'^{m}u'^{n}+s'^{n}u'^{m}]\nonumber\\&&\times
\bigg[(\xi_1.\xi_2)\frac{1}{2}u's'
\bigg]\bigg).\labell{amphigh}\eeqa

In order to check these proposed infinite couplings \reef{lagrango}, we are going to compare \reef{amphigh} with the infinite number of the tachyon poles in  string theory for several values of $n,m$. Note that we have removed the common factors in both string and field theory sides.

First we set, $n=m=0$, then \reef{amphigh} gives us a coefficient as :

\beqa
-8(a_{0,0}+b_{0,0})&=&-8(\frac{-\pi^2}{6}+\frac{-\pi^2}{12})=2\pi^2\nonumber\eeqa

On the other  hand, we have  a factor of  $(2\pi^2c_{0,0})$ in \reef{1pptc}. Comparing  $(2\pi^2c_{0,0})$ with the factor of $(2\pi^2)$ in field theory one gets the consistent result. Let us proceed at first order of  $\alpha'$, so \reef{amphigh} does include a factor as
\beqa
-4(a_{1,0}+a_{0,1}+b_{1,0}+b_{0,1})(s'+u')&=&0\nonumber\eeqa

  Equation \reef{1pptc} is also proportional to   $\pi^2(c_{1,0}+c_{0,1})(s'+u')$, which is indeed zero.  At second order or $(\alpha')^2$, \reef{amphigh} does involve a coefficient as
\beqa
&&-8(a_{1,1}+b_{1,1})s'u'-4(a_{0,2}+a_{2,0}+b_{0,2}+b_{2,0})[s'^2+u'^2]\nonumber\\
&&=\frac{\pi^4}{3}(2s'u')+\frac{2\pi^4}{3}(s'^2+u'^2)
\nonumber\eeqa
Again \reef{1pptc} does have a factor  $\pi^2[c_{1,1}(2s'u')+(c_{2,0}+c_{0,2})(s'^2+u'^2)]$, which is precisely equivalent to string amplitude ( by making use of \reef{577}).
In order  $\alpha'^3$, \reef{amphigh} has the following coefficient
\beqa
&&-4(a_{3,0}+a_{0,3}+b_{0,3}+b_{3,0})[s'^3+u'^3]-4(a_{1,2}+a_{2,1}+b_{1,2}+b_{2,1})[s'u'(s'+u')]\nonumber\\
&&=-16\pi^2\xi(3)s'u'(s'+u')
\nonumber\eeqa
and in \reef{1pptc} we have the following factor $\pi^2[(c_{0,3}+c_{3,0})[s'^3+u'^3]+(c_{2,1}+c_{1,2})s'u'(s'+u')]$.  
Finally at order  $(\alpha')^4$, \reef{amphigh} consists a factor of
\beqa
&&-4(a_{4,0}+a_{0,4}+b_{0,4}+b_{4,0})(s'^4+u'^4)-4(a_{3,1}+a_{1,3}+b_{3,1}+b_{1,3})[s'u'(s'^2+u'^2)]\nonumber\\
&&-8(a_{2,2}+b_{2,2}) s'^2u'^2=\frac{4\pi^6}{15}(s'^4+u'^4+2(s'^3u'+u'^3s')+3s'^2u'^2)
\nonumber\eeqa
and \reef{1pptc} shows  the factor of $\pi^2[(c_{4,0}+c_{0,4})(s'^4+u'^4)+(c_{1,3}+c_{3,1})(s'^3u'+u'^3s')+2c_{2,2} s'^2u'^2]$ which is surprisingly equal to the above factor making use of  the coefficients in \reef{577}.
One might conclude that these comparisons can be easily extended\footnote{Similar computations for the amplitude of one RR and three massless open strings have been checked in \cite{Hatefi:2010ik,Hatefi:2012ve,Hatefi:2012rx}.} to all orders of $\alpha'$. Thus, \reef{amphigh} precisely does reproduce the infinite  number of tachyon poles of the string amplitude of \reef{1pptc}.

This shows  that in addition to higher derivative couplings of two scalars and two tachyons are
being exact up to on-shell ambiguity, the momentum expansion of the amplitude $C\phi\phi T$ does agree  with $TT\phi\phi$'s  momentum expansion .

\section{$p=n+1$ case}

  The trace in ${\cal A}_2$ amplitude can be done as follows
\beqa
\Tr\bigg(\fsH_{(n)}M_p
\Gamma^{ba}\bigg)&=&\pm\frac{32}{n!}\eps^{a_{0}\cdots a_{p-2} ba}H_{a_{0}\cdots a_{p-2}}
,\nonumber\eeqa

Taken into account this trace in ${\cal A}_2$,  one gets  the amplitude in string side as
\beqa
{\cal A}^{\phi\phi TC}&=&\mp\frac{32i}{(p-1)!}(\mu'_p\beta'\pi^{1/2})\Tr(\lam_1\lam_2\lam_3)H_{a_{0}\cdots a_{p-2}}\eps^{a_{0}\cdots a_{p-2}ba}\bigg\{
2k_{1a}k_{2b}\xi_1.\xi_2 L_2
 \bigg\}\labell{line97}.\eeqa

 As it is obvious from \reef{line97} the amplitude is antisymmetric with respect to two scalar fields. It shows that the amplitude must have zero value just for the Abelian gauge.  The amplitude also does have an infinite number  of massless poles in the t-channel and an infinite number of  contact interactions. First we produce all infinite numbers of  massless poles and then we come to all contact interactions. Now we want to make an important comment.  Because of some kinematic reason in our amplitude we do not have any tachyon pole and this is unlike the scattering amplitude of two gauge fields, one RR and one tachyon so we do not have any tachyon pole in  field theory  either.

 \subsection{Infinite number of massless poles and contact interactions for $p=n+1$ case}

Having replaced  the expansion of $L_2$ and the related trace  into ${\cal A}_2$, one comes to an  infinite number of the massless poles in the $t$-channel as
\beqa
{\cal A}^{\phi\phi TC}&=&\pm\frac{32i\mu'_p\beta'\pi^{2}}{t(p-1)!}\Tr(\lam_1\lam_2\lam_3)H_{a_{0}\cdots a_{p-2}}
\eps^{a_{0}\cdots a_{p}ba}\sum_{n=-1}^{\infty}b_n(u+s+1/2)^{n+1}\nonumber\\&&\times
\bigg[2\xi_1.\xi_2k_{1a}k_{2b}
\bigg],
\labell{masspole}\eeqa

There are also an infinite
 number of contact interactions for this case that we want to consider  later on.  In field theory, these massless poles should be read off
with this   Feynman diagram
\beqa
{\cal A}&=&V_\alpha^{i}(C_{p-2},T_3,A)G_{\alpha\beta}^{ij}(A)V_\beta^{j}(A,\phi_1,\phi_2),\labell{amp54}\eeqa

As observed from the expansion of the amplitude, there are an infinite number of higher derivative couplings between one RR (p-2)-form field ($C_{p-2}$), one on-shell tachyon and one off-shell gauge field where they are related to the higher derivative corrections of the WZ coupling $\Tr(C_{p-2}\wedge F\wedge DT)$. They have been found in \cite{Garousi:2008ge}. Therefore the vertex of $V_\alpha^{i}(C_{p-2},T_3,A)$  must be found from  the higher derivative corrections of this WZ coupling   such that gauge field has to be off-shell. Thus the needed vertices and gauge field propagator are given as
\beqa
V_\alpha^{i}(C_{p-2},T_3,A)&=&2\mu'_p\beta'(2\pi\alpha')^{2}\frac{1}{(p-1)!} \eps^{a_{0}\cdots a_{p-1}i}H_{a_{0}\cdots a_{p-2}}k_{a_{p-1}}\sum_{n=-1}^{\infty}b_n(\alpha'k_3\cdot k)^{n+1}\Tr(\lam_3\Lambda^{\alpha}),\nonumber\\
V_\beta^{j}(A,\phi_1,\phi_2)&=&-iT_p(2\pi\alpha')^{2} \xi_{1}.\xi_{2}(k_1-k_2)^{j}[-\Tr(\lambda_{1}\lambda_{2}\Lambda_{\beta})+\Tr(\lambda_{2}\lambda_{1}\Lambda_{\beta})],
\label{mmnn1}\\
G_{\alpha\beta}^{ij}(A)&=&\frac{i\delta_{\alpha\beta}\delta^{ij}}{(2\pi\alpha')^{2}T_p(t)}\nonumber\eeqa

where the gauge field propagator must be found  from its kinetic term, which is given in  \reef{expandL2}.
On the other hand, the vertex of  $V_\beta^{j}(A,\phi_1,\phi_2)$ has been reduced from  the scalar field's kinetic term  $\frac{\lambda^2}{2}  \Tr(D_a\phi_i D^a\phi^i)$. The other remark that should be made is that, in order to produce the desired amplitude for 123 ordering we must keep just the first term in the vertex of $V_\beta^{j}(A,\phi_1,\phi_2)$. As always  $k$ is  the off-shell gauge field's momentum. Regarding the point that  massless poles of amplitude have no higher derivative corrections, one may understand that  scalar field's kinetic term does not involve correction so
the vertex $V_\beta^{j}(A,\phi_1,\phi_2)$ does not have a higher derivative correction either, because  it has already been fixed in the DBI action.

One replaces \reef{mmnn1} into  \reef{amp54}, to get
\beqa
{\cal A}&=&(2\pi\alpha')^{2}\frac{2\mu'_p\beta'}{(p-1)!t}\eps^{a_{0}\cdots a_{p-1}j}H_{a_{0}\cdots a_{p-2}}
\Tr(\lambda_{1}\lambda_{2}\lambda_{3})\sum_{n=-1}^{\infty}b_n\bigg(\frac{\alpha'}{2}\bigg)^{n+1}(s+u+1/2)^{n+1}
\nonumber\\&&\times\bigg(-2(\xi_1.\xi_2)k_{2a_{p-1}}k_{1j}\bigg).\labell{ver22}\eeqa

which is precisely the infinite number of the t-channel massless poles of  \reef{masspole}.
As a comment and unlike $p=n-1$ case, here there should not be any residual contact interactions in \reef{ver22}. Indeed we get this result
after comparing  all  massless poles of field theory  \reef{ver22} with the infinite number of massless poles in string side.

 This does show  that the momentum expansion of our amplitude $C\phi\phi T$  is actually consistent, even with the momentum expansion of $CA T$.
Let us end this section by producing all infinite numbers of contact terms for $p=n+1$ case in field theory  .

Having replaced $L_2$ into the S-matrix in ${\cal A}_2$, one gets contact terms at all orders of $\alpha'$ in string theory
as
\beqa
{\cal A}^{\phi\phi TC}&=&\mp\frac{32i}{(p-1)!}(\mu'_p\beta'\pi^{2})\Tr(\lam_1\lam_2\lam_3)H_{a_{0}\cdots a_{p-2}}\eps^{a_{0}\cdots a_{p-2}ba}\nonumber\\&&\times\bigg(
2k_{1a}k_{2b}\xi_1.\xi_2\bigg) \bigg(\sum_{p,n,m=0}^{\infty}e_{p,n,m}t^{p}(s'u')^{n}(s'+u')^m\bigg).
\labell{line8997}\eeqa

These infinite numbers of contact terms in \reef{line8997} can be reproduced by taking the following gauge invariant couplings:

\beqa
S_{1}&=&2i\lambda^2\beta'\mu'_p\int d^{p+1}\sigma {1\over (p-2)!}(\veps^v)^{a_0\cdots a_{p}}\sum_{p,n,m=0}^{\infty}
e_{p,n,m}\left(\alpha'\right)^{2n+m+1}\left(\frac{\alpha'}{2}\right)^{p}
\nonumber\\&&\times C^{(p-2)}_{a_0\cdots a_{p-3}}(\sigma)\Tr\left( D^{a_1}\cdots D^{a_{2n}} D^{b_1}\cdots D^{b_{m}}D_{a_{p-2}}T\right.\nonumber\\
&&\left. (D^aD_a)^p  D_{b_1}\cdots D_{b_{m}}\bigg[D_{a_1}\cdots D_{a_n}D_{a_{p-1}}\phi^i D_{a_{n+1}}\cdots D_{a_{2n}}D_{a_{p}}\phi_i\bigg]\bigg)
\right.
.\labell{hderv1989}
\eeqa

  Here $(\veps^v)$ is the volume form which takes place in subspace parallel to the brane's world volume.
 Note that, to produce \reef{line8997},  one may be able to write down another higher derivative gauge invariant couplings, to make a contribution to the contact terms of $C\phi\phi T$ amplitude. Thus \reef{hderv1989} is in fact one special higher derivative gauge invariant coupling  that  results all terms in \reef{line8997}.

\vskip 0.2in

So we get to an important remark. The leading order terms of our amplitude related to Wess-Zumino couplings and naturally  higher order terms must correspond to WZ couplings' higher derivatives .

Therefore we learned that by analyzing massless poles we get useful information about the higher derivative corrections of $\Tr(C_{p-2}\wedge F\wedge DT)$. By studying contact interactions we gain remarkable information on new coupling as   $\Tr(C_{p-2}\wedge DT\wedge D\phi\wedge D\phi)$.

\vskip 0.1in

It is worth  talking about some details related to  WZ couplings. They  can be derived also with BSFT method. However as noted in   \cite{Kraus:2000nj} setting constant RR field reduces to having no higher derivative correction to these WZ couplings.

 Notice the point that we have already mentioned. Our derived couplings make sense in the presence of $p_ap^a\rightarrow\frac{1}{4}$.  So
 the conclusion as a matter of fact is that we are not allowed to compare our couplings with the constant RR field as a
 ($p_ap^a=0$) result of  the BSFT.

\subsection{Contact terms}

Doing in detail all an infinite number of tachyon and massless poles of string theory amplitude \reef{44}, we are now ready to
extract the rest of the contact terms of the amplitude. As can be seen from the poles of the Gamma function,   $(-t-s'-u')L_1$  has neither a massless pole nor a tachyon pole. We will show now this consistency with the WZ terms . By setting $n=p+1$ to all terms in $ A_{1}$ in equation \reef{44} except the last term, the string scattering amplitude
takes the  following form:
\beqa
A_{1}^{C\phi\phi T}&=&\frac{32i\pi^3\mu'_p\beta'}{(p+1)!}\xi_1^i\xi_2^j\left(
-p_{i}p_{j}(H^{(p+1)})_{a_0\cdots a_p}
\right. \nonumber\\
&&\qquad\qquad\qquad
+(p+1)k_{1a_0}p_{j}(H^{(p+1)})_{ia_1\cdots a_p}+
(p+1)k_{2a_0}p_{i}(H^{(p+1)})_{ja_1\cdots a_p}
\nonumber\\
&&\left.\qquad\qquad\qquad+p(p+1)\,k_{1a_0}k_{2a_1}
(H^{(p+1)})_{ija_2\cdots a_p}\right)
(\veps^v)^{a_0\cdots a_p}
\nonumber\\&&
\times\bigg( \sum_{n=0}^{\infty}c_n(s'+t+u')^{n+1}+\sum_{n,m=0}^{\infty}c_{n,m}[(s')^n(u')^m +(s')^m(u')^n]\nonumber\\&&
+\sum_{p,n,m=0}^{\infty}f_{p,n,m}(s'+t+u')^{p+1}[(s'+u')^n(s'u')^{m}]\bigg),
\labell{452o}
\eeqa
where $H^{(p+1)}=dC^{(p)}$.
In fact, by analyzing the $C\phi\phi T$ amplitude, one understands that it needs some interaction terms in which either must come from
 pullback or Taylor expansion of one RR p-form field ($C^{(p)}$).
Due to not having any external gauge field here we just replaced all covariant derivative of scalars and tachyon  with their partial derivatives.

One may check  that the leading contact terms  in our amplitude can be reproduced  by considering the following field
interactions:
\beqa
S_{2}&=&\frac{\lambda^3\beta'\mu'_p}{2p!}\int d^{p+1}\sigma\,(\veps^v)^{a_0\cdots a_p}
\left( Tr(\partial_{a_0}T
\Phi^i\Phi^j)\,\prt_i\prt_jC^{(p)}_{a_1\cdots a_p}\right.\nonumber\\
&&\qquad\qquad\qquad
+2p\,\Tr(\prt_{a_0}T\prt_{a_1}\Phi^i\Phi^j)\,\prt_j C^{(p)}_{ia_2\cdots a_p}
\nonumber\\
&&\qquad\qquad\qquad\left.+\frac{}{}(p-1)p\,\Tr(\partial_{a_0}T\prt_{a_1}
\Phi^i\prt_{a_2}\Phi^j)\,C^{(p)}_{ija_3\cdots a_p}\right),
\labell{s22}
\eeqa

Applying integration by parts,
 these contributions \reef{s22} might be reconsidered as
\beqa
S_{2}&=&\frac{\lambda^3\beta'\mu'_p}{2p!}\int d^{p+1}\sigma\,(\veps^v)^{a_0\cdots a_p}
\left( Tr(\partial_{a_0}T
\Phi^i\Phi^j)\,\prt_j H^{(p+1)}_{ia_1\cdots a_p}\right.\nonumber\\
&&\qquad\qquad\qquad
+ p\,\Tr(\partial_{a_0}T\prt_{a_1}\Phi^i\Phi^j)\, H^{(p+1)}_{ija_2\cdots a_p}\bigg),
\labell{s227i}
\eeqa

where for higher derivative corrections our notation is as follows:
\beqa
(s'u')^{m}HT\phi\phi&=&(\alpha')^{2m}H D_{a_1}\cdots D_{a_{2m}}T\partial^{a_{1}}\cdots \partial^{a_{m}}\phi
\partial^{a_{m+1}}\cdots \partial^{a_{2m}}\phi,\nonumber\\
(s'+u')^{n}HT\phi\phi&=&(\alpha')^{n}H D_{a_1}\cdots D_{a_{n}}T\partial^{a_{1}}\cdots \partial^{a_{n}}(\phi
\phi),\nonumber\\
(s')^{n}u'^m HT\phi\phi&=&(\alpha')^{n+m}H D_{a_1}\cdots D_{a_{n}} D_{a_{1}}\cdots D_{a_{m}}T \partial^{a_{1}}\cdots \partial^{a_{n}}\phi\partial^{a_{1}}\cdots \partial^{a_{m}}\phi,
\nonumber\\
(s'+t+u')^{p+1} HT\phi\phi&=&(\frac{\alpha'}{2})^{p+1}H (D_{a}D^{a})^{p+1}(T\phi\phi).
\labell{67s282ee}
\eeqa
Now we write down the contact terms that are related to the last term of $A_1$ in \reef{44}.
Note that it is not possible to have a massless scalar field as an off-shell state.
Because if we imagine there was a nonvanishing coupling between one  scalar field and one RR (p+1)-form field strength
  then the interaction in WZ action would be

\beqa
\lambda\mu'_p\int d^{p+1}\sigma {1\over (p+1)!}
(\veps^v)^{a_0\cdots a_{p}}\,\Tr\left(\Phi^i\right)\,
H^{(p+2)}_{ia_0\cdots a_{p}}(\sigma)\ \nonumber\eeqa

Therefore this off-shell scalar created by that interaction
should be attached within  an interaction including two on-shell scalars and
one on-shell tachyon. However, there are no world-volume interactions between three scalars and one tachyon. The same thing happens for the case of an off-shell
gauge field because it will not have any coupling with the RR (p+1)-form field.
Thus the amplitude for $n=p+1$ does not involve any scalar/gauge pole.

\vskip 0.1in

Extracting the trace and keeping just the contact terms for the last term in $A_{1}$  we end up with the following terms
\beqa
\label{l1cts}
A_{1}^{c\phi\phi T}&=&\frac{16i\pi^3\mu'_p\beta'}{(p+1)!}\xi_1.\xi_2\left(
(H^{(p+1)})_{a_0\cdots a_p}(\veps^v)^{a_0\cdots a_p}\right)
\bigg( \sum_{n=0}^{\infty}c_n(s'+t+u')^{n}s'u'\nonumber\\&&
+\sum_{p,n,m=0}^{\infty}f_{p,n,m}(s'+t+u')^{p}[(s'+u')^n(s'u')^{m+1}]\bigg),
\labell{p=n case32}\eeqa

The above contact terms in this part of the amplitude  are now a sort of  new couplings which can be produced by the following field interactions in field theory  as well:

\beqa
S_{3}&=&\frac{i\lambda^3\beta'\mu'_p}{4p!}\int d^{p+1}\sigma\,(\veps^v)^{a_0\cdots a_p}
\left( Tr(\partial_{a_0}T
\Phi^i\Phi_i)\,C^{(p)}_{a_1\cdots a_p}\right),
\labell{snjh22}
\eeqa
Contact terms in the first line of  \reef{p=n case32} will be reproduced by taking the following couplings:
\beqa
-\lambda^3\beta'\mu'_p \sum_{n=0}^{\infty}c_n\left(\frac{\alpha'}{2}\right)^n C_{p}\wedge (D^aD_a)^n[D^aD^bDT(D_a\phi^i D_b\phi_i)],\labell{hderv6775}
\eeqa
It becomes so clear from \reef{hderv6775} that the nonleading order  terms do correspond  to the higher derivative corrections  of \reef{snjh22}.
Indeed it looks like the coupling which we found in non-BPS formalism $C_p\wedge DT \,T^2$.
The contact terms in the second line of \reef{p=n case32} might correspond to a coupling such as
\beqa
&&-\lambda^3\beta'\mu'_p
\sum_{p,n,m=0}^{\infty}f_{p,n,m}(\alpha')^{2m+n}\left(\frac{\alpha'}{2}\right)^pH_{p+1}(D_aD^a)^p \labell{highhigh}\\
&& \left[D_a D_b D^{a_1}\cdots D^{a_n}D_{b_1}\cdots D_{b_{2m}}TD_{a_1}\cdots D_{a_n}(D^aD^{b_1}\cdots D^{b_m} \phi^i D^b D^{b_{m+1}}\cdots D^{b_{2m}}\phi_i) \right].\nonumber\eeqa

These new interactions in field theory  are  completely  consistent with the string theory amplitude of one RR $p$- form field, two scalar fields and one tachyon.  Therefore,
it shows up that perturbative string theory is a strong tool to discover new couplings in field theory side.

Eventually let us come back to  the contact interaction terms that amplitude \reef{amp544u} have resulted. Applying some identities, we write down all contact interactions as:

\beqa
32i\pi\beta'\mu_p'\frac{ \eps^{a_{0}\cdots a_{p}}H_{a_{0}\cdots a_{p}}}{(p+1)!}\bigg[(\xi_1.\xi_2)\frac{1}{2}u's'
\bigg]
\sum_{n,m=0}^{\infty}(a_{n,m}+b_{n,m})(-\alpha' k^2-\alpha' m^2)^{l-1}\nonumber\\
\left[\left(4\sum_{l=1}^m \pmatrix{m\cr
l}(s'^{m-l}u'^n+u'^{m-l}s'^n)+4\sum_{l=1}^n \pmatrix{n\cr
l}(s'^{n-l}u'^m+u'^{n-l}s'^m)\right) \right.\nonumber\\
\left.+\sum_{l=1,j=1}^{n,m} \pmatrix{n\cr
l}\pmatrix{m\cr j}(s'^{n-l}u'^{m-j}+u'^{n-l}s'^{m-j})(-\alpha'k^2-\alpha' m^2)^{j}\right]
\Tr(\lam_1\lam_2\lam_3)\nonumber\eeqa
Notice that these interactions could be reconsidered  in the following form:
\beqa
32i\pi\beta'\mu_p'\frac{ \eps^{a_{0}\cdots a_{p}}H_{a_{0}\cdots a_{p}}}{(p+1)!}\bigg[(\xi_1.\xi_2)\frac{1}{2}u's'
\bigg]\Tr(\lam_1\lam_2\lam_3)
\sum_{p,n,m=0}^{\infty}f'_{p,n,m}(s'+t+u')^p(s'+u')^n(s'u')^{m},\label{kkoo}\eeqa

One may be able to write $f'_{p,n,m}$  in terms of $a_{n,m}$ and $b_{n,m}$. One has to consider the fact that the last contact terms in   \reef{high6587} have the same structure as those terms appeared in \reef{kkoo}. Thus the coefficients $f_{p,n,m}$ in \reef{highhigh}  must be substituted with \beqa
f_{p,n,m}\rightarrow f_{p,n,m}-f'_{p,n,m}\nonumber\eeqa
Therefore the higher derivative  theory  will exactly give rise to the string theory  amplitude.
\subsection{Contact terms for  $n=p+3$}

For this case which includes a  RR
(p+2)-form field , one gets the fact that  exchanging a massless gauge/scalar is not allowed.  In particular, the (p+2)-form potential
has one rank more than RR (p+1)-form potential to result in a desired interaction in world  volume space.
Thus for this case we might expect that our amplitude does not involve any massless/tachyon
poles. Thus the leading terms of the  amplitude are in fact just contact terms that we want to identify now.
The infinite number of contact terms in string amplitude is given as
\beqa
{\cal A}_{3}&=&  \frac{32i\pi^2\mu'_p\beta'}{(p+3)!}(H^{(p+3)})_{ija_0\cdots a_p}\xi_{1i}\xi_{2j}(\veps^v)^{a_0\cdots a_p} \nonumber\\&&
\times
\bigg(\sum_{n=-1}^{\infty}b_n(u'+s')^{n+1}+\sum_{p,n,m=0}^{\infty}e_{p,n,m}t^{p+1}(s'u')^{n}(s'+u')^m\bigg).\labell{high67}\eeqa

In order to reproduce all terms in \reef{high67} we begin with the WZ action. The minimal interaction for this case does include
a RR (p+2)-form field in the bulk, two  scalars within commutator and one on-shell tachyon. Thus
the interaction that does include the RR (p+2)-form potential is

\beqa
S_{4}&=&\frac{(2\pi\alpha') \lambda\beta'\mu'_p}{p!}\int d^{p+1}\sigma\,(\veps^v)^{a_0\cdots a_p}
\left( Tr(\partial_{a_0}T
[\Phi^i,\Phi^j])\right)C^{(p+2)}_{jia_1\cdots a_p},
\labell{s89h22}
\eeqa
 so by extracting commutator we get  the following leading nonvanishing coupling which is confirmed by direct S-matrix computations as follows:
\beqa
S_{4}&=&\frac{(2\pi\alpha') 2\lambda\beta'\mu'_p}{p!}\int d^{p+1}\sigma\,(\veps^v)^{a_0\cdots a_p}
\left( Tr(\partial_{a_0}T
\Phi^i\Phi^j)\right)C^{(p+2)}_{jia_1\cdots a_p},
\labell{s89h22}
\eeqa
where for higher derivative corrections our notation is as follows:
\beqa
(t)^{p} HT\phi\phi&=&(\frac{\alpha'}{2})^{p}H T(D_{a}D^{a})^{p}(\phi\phi).
\labell{67s282}
\eeqa

So we have produced  the higher derivative  theory  with  the string theory  amplitude of $C\phi\phi T$. It is now fair to say that we have shown the complete consistency of one tachyon, two scalars and one closed string RR around  \reef{point} with all its higher derivative couplings in field theory.

This ends our goal which was to show complete consistency between string theory scattering amplitudes for different values of $p,n$  and making use of symmetrized trace tachyon DBI action.

\section{Concluding remarks}

 As it is seen for the simple tachyon poles, the coupling of  $\Tr(C_{p}\wedge DT)$ does not get any correction. Therefore by studying nonleading tachyon poles one can find  information about the higher derivative corrections to the coupling of two scalar fields and two tachyons  where we found them up to all orders of $\alpha'$ in \reef{lagrango}. This is worth mentioning that  contact terms of the string amplitude already include   information about the higher derivative corrections to $\Tr(C_{p-2}\wedge DT \wedge D\phi^i\wedge D\phi_i)$, $\Tr(C_{p}\wedge DT \phi^i\phi_i)$ and
 $
 \Tr(\partial_{a_0}T
[\Phi^i,\Phi^j])C^{(p+2)}_{jia_1\cdots a_p}(\veps^v)^{a_0\cdots a_p}$. Note that here we obtained a new  coupling between the commutator of scalar fields with covariant derivative of tachyon and one RR.
  It is an interesting issue to study the scattering amplitude  between one RR, one tachyon, one gauge field and one scalar field in the world volume of non-BPS branes to get more information on WZ couplings. It would be also nice to follow this amplitude in order to get all the information needed about the higher derivative corrections of this amplitude to WZ  and tachyonic actions \cite{Hatefi:2012eh}.

\subsection {Discussion}

We have done the S-matrix elements of $CT\phi\phi$ in the world volume of a single non-BPS SD-brane. Having set the on-shell conditions, we believe that $CT\phi\phi$ amplitude makes sense just in the presence of a single non-BPS SD-brane. Applying its momentum expansion, we were able to show the consistency of the leading order terms of the expansion with the WZ couplings of a single non-BPS SD-brane. The non-leading terms have been extracted with the help of  some higher derivative corrections of the WZ couplings. In fact, they have been produced with the  equations \reef{hderv1989}, \reef{s227i}, \reef{67s282ee} , \reef{hderv6775} and \reef{highhigh}.

$\frac{}{}$

 The amplitude  of $C\phi\phi T$ consists of two parts.  The first part does include a RR $p$-form ($C_p$) with an infinite number of tachyon poles and  many contact interaction terms. The contact interactions  give rise to a new coupling of the form $ \Tr(C_p\wedge DT\phi^i\phi_i) $ and give some higher derivative corrections to this coupling. They are in precise agreement with S-matrix elements of this amplitude  for $p+1=n$ case.  To reproduce an infinite number of tachyon poles, one has to find  two tachyon and  two scalar field couplings  to all orders of $ \alpha'$ where we were able to find them in \reef{lagrango} with direct S-matrix computations. We could  confirm and check them with tachyon poles in our S-matrix computations.
\vskip 0.1in

Note also that comparing an infinite number of field theory tachyon poles with the infinite poles of string theory  gave rise to some residual contact interactions. Due to having  the same structure for these contact interactions as those that appeared in the second line of \reef{high6587}, we should be able to modify the coefficients $f_{p,n,m}$ in \reef{highhigh}. Therefore one must substitute $f_{p,n,m}-f'_{p,n,m}$  instead of $f_{p,n,m}$ in \reef{highhigh}.

$\frac{}{}$

 The second part does contain a RR (p-2)-form field ($C_{p-2}$) with an infinite number of massless poles and many contact terms. All massless poles must be reproduced by the higher derivative couplings of one RR $(p-2)$-form field, one Abelian field strength and covariant derivative of the tachyon or in the other words with the coupling of $\Tr(C_{p-2}\wedge DT\wedge F)$  and by making use of kinetic term of scalar fields. One may point out that in this part of the amplitude we have to use the commutator of the scalar and gauge field as well. In order to reproduce contact terms of  $C\phi\phi T$  for  $p=n+1$ case, the coupling  of $\Tr(C_{p-2}\wedge DT\wedge D\phi_i\wedge D\phi^i)$ and its higher derivative corrections must be taken into account. We have also fixed the coefficient of this coupling.
\vskip 0.1in
 In the present paper, making use of direct string computations, we have confirmed the existence of the commutator of scalars $[\Phi^i,\Phi^j]$
which could come  from
either the exponential inside of the WZ action, or the expansion of the det($Q$) in the non-Abelian DBI action, where for this amplitude they come from the exponential within the WZ action.
\vskip 0.1in

Indeed  to produce all infinite numbers of massless poles of the amplitude for $p=n+1$ case
we had to use commutator of guage and scalar in order to have the vertex of two on-shell scalars and one off-shell gauge field where the contribution is coming from
 the commutator in the definition of $D_a\Phi^i$ .

\vskip 0.1in

It is really important to highlight the point that we could find out a very interesting coupling between the commutator of two scalar fields and the partial derivative of the tachyon and one RR  as the following:
\beqa
S_{4}&=&\frac{(2\pi\alpha') \lambda\beta'\mu'_p}{p!}\int d^{p+1}\sigma\,(\veps^v)^{a_0\cdots a_p}
\left( Tr(\partial_{a_0}T
[\Phi^i,\Phi^j])\right)C^{(p+2)}_{jia_1\cdots a_p},
\nonumber
\eeqa

Of course in this amplitude we just had one external tachyon and two scalars and one RR, that is why we drop the commutator between tachyon and gauge field in the above coupling in the definition of $DT$ and just considered the partial derivative of tachyon. We may suppose that in the above coupling perhaps there should be a  covariant derivative of the tachyon instead of its partial derivative. It would be nice to confirm  this coupling by doing a six-point function $CTA\phi\phi$, which we leave it for future work \cite {Hatefi:2012eh1}.

\vskip 0.1in

One more remarkable thing is that these potential interactions do depend strongly on the field
strength of the RR. Thus the gauge invariance for the RR must be held as well. However,
as it is clear from the Wess-Zumino action \reef{finalcs}, all interaction terms were written
 in terms of RR potentials  or just $C$ terms so one might realize that the RR gauge
invariance has not been satisfied any more. However, for the interactions appeared in this paper, we were able to show that both
representations for RR couplings do agree and the difference between them is some total derivative terms.

\vskip 0.1in

Another important comment should be made is as follows. Having taken integration by parts we arrive at some
complicated couplings between terms which indeed have different
notions. For example the terms coming from the interior product inside  WZ action or the terms coming from the Taylor expansion or the the terms coming from pullback are completely mixed in applying the expansion of the Wess-Zumino action.

\vskip 0.1in


It is shown that  the  expansion of the amplitude of one RR, one tachyon  and two scalar fields $C\phi\phi T$ around   \reef{point} does belong  to the higher derivative correction of the Wess-Zumino terms. Therefore  we truly  believe  that  \reef{point} is a unique momentum expansion of four-point functions including one RR in the bulk $(C)$, one tachyon $(T)$ and either two scalar fields $ (\phi\phi)$ or two gauge fields $(AA)$. Notice that the leading order term in the amplitude does correspond to the mentioned effective actions and all nonleading terms of the amplitude are consistent with  the higher derivative corrections of the effective actions.  On general grounds we realize that this expansion also holds for $CTA\phi$ amplitude \cite{Hatefi:2012eh1}.

\vskip 0.1in

Let us compare \reef{point} with  the momentum expansion of the S-matrix elements including three massless open strings \cite{Hatefi:2010ik,Hatefi:2012ve,Hatefi:2012rx} and one RR closed string. Momentum expansion for these cases  has been found \cite{Hatefi:2010ik}, which is  $\alpha' k_i\cdot k_j\rightarrow 0$. In fact it is equal to $\alpha'(k_i+k_j)^2\rightarrow 0$. While, if we include tachyon to the amplitude then we can conclude that the expansion around $\alpha' k_i\cdot k_j\rightarrow 0$ for the tachyon for sure is not the same as  $\alpha'(k_i+k_j)^2\rightarrow 0$.
\\

One important facet of these couplings is that  they do include  on-shell ambiguity. To resolve it, one has to compare it with some off-shell interactions which already appeared in BSFT formalism. It was seen in \cite{Kraus:2000nj} that the WZ couplings are being exact once one considers the RR field as a constant field. Therefore, we have to write  all interactions  in the momentum space in terms of the Mandelstam variables  such that now they will be written  in terms of Ramond-Ramond's momentum. The Mandelstam variables for the amplitude  of $CT\phi\phi$ must be sent to $t\rightarrow 0$, $s\rightarrow -1/4$ and $u\rightarrow -1/4$. Now, one may be able to reconsider them as
$(p^2+2k_1\cdot p)\rightarrow 0,(p^2+2k_2\cdot p)\rightarrow 0$ and $(p^2+2k_3\cdot p)\rightarrow \frac{-1}{4}$. Applying a constant RR field for these forms of the couplings one immediately concludes that there are no contributions for all higher derivative corrections.

\vskip 0.1in

 However, once more remember that for non-BPS SD-branes,  those couplings  make sense just in the appearance of an extra assumption which is  $p_ap^a\rightarrow 1/4$.   Thus we are not allowed to compare our higher derivative couplings with the interactions coming from BSFT.

\vskip 0.1in

Note that the  couplings   $TT\phi\phi$ in \reef{lagrango} might have on-shell ambiguity which means that by replacing $m^2T$ with $ DDT$ tachyon poles will not change but rather would produce some more additional contact interactions. Thus by performing the amplitude of two tachyons, two scalar fields and one gauge field  in which couplings \reef{lagrango} would appear in tachyon poles and in contact interactions of this amplitude, we might solve all those  ambiguities. It would be interesting
to carry out this rather long but straightforward computation \cite{Hatefi:2012eh}.

\vskip 0.2in

Let us address some more unsolved problems, where one of them has been already addressed in \cite{Garousi:2008tn}.

Therefore an interesting amplitude to solve in favor of symmetrized trace is the scattering amplitude of $CTTTT$. In  order to investigate whether symmetric trace works for  tachyon DBI action or ordinary trace, one must study in detail the S-matrix element of one closed string Ramond-Ramond and four tachyons in  the system of brane-antibrane in which the infinite number of the simple tachyon poles have to result in  the  following Feynman diagram
\beqa
 A&=&V^{\alpha}(C_{P-1},T,T)  G_{\alpha\beta}(T)V^{\beta}(T,T,T,T)\nonumber\eeqa
 \vskip 0.2in

This amplitude will solve this apparent ambiguity for the tachyon action. In favor of applying symmetric trace it is really a good test to  follow this computation within detail.
\vskip 0.1in

 As we have already observed extracting the higher order contact interactions
is really extremely tedious. We have made some progress in finding
full consistency of the scattering amplitudes including one closed string RR, one tachyon and two scalar fields in the world volume of a single non-BPS SD-brane.
 It is evident from other investigations
that commutator terms play the key  role at higher orders.
It would be nice to check the details of some six-point functions in order to remove some ambiguities \cite{Hatefi:2012eh44}.

\vskip 0.1in

 Symmetrized trace was informed in \cite{Tseytlin:1997csa} to describe low energy gauge theory
 with some simple background fields. It was  shown that this kind of trace prescription does need
 corrections
at sixth order in gauge field's field strength \cite{Hashimoto:1997gm}. In non-Abelian gauge theory these problems  have been addressed in terms of the
ambiguity between interchanging  field strengths and covariant derivatives . One imagines that the  commutators
of field strengths might be redefined in terms of their covariant
derivatives. Thus it would be interesting to discover  within detail some six point functions to find
interactions including some high derivatives of the gauge field's field strengths\cite{Hatefi:2012eh44}. Some progresses have been made \cite{Cornalba:2000wq}, also through applying ideas of non commutative field theory  some investigations had been done  \cite{Douglas:1997fm}.
\\
$\frac{}{}$
 \\
Let us address the final point. In order to get more information in finding higher derivative corrections for both tachyonic DBI  and Wess-Zumino effective actions it is really a very nice issue to pursue the amplitude of one closed string RR field, one gauge field, one scalar field and one tachyon in the world volume of  non-BPS branes \cite{Hatefi:2012eh}.





\section*{Acknowledgments}
The author wishes to thank K.S.Narain, F.Quevedo, N.Arkani-Hamed and I.Y. Park for helpful conversations. He would also like to thank J.Maldacena for comments and E.Martinec for useful suggestions.  He also acknowledges L. \'Alvarez-Gaum\'e , G.Veneziano, N.Lambert and A.Sagnotti for several valuable discussions.
\section{Appendices}
\renewcommand{\thesection}{Appendix }
\setcounter{equation}{0}
\renewcommand{\theequation}{ \arabic{equation}}
\appendix
\section{Appendix A : Doubling trick, some useful correlation functions}\label{integrals}


 In order to deal with standard holomorphic conformal field theory propagators on the boundary of world sheet we might use the doubling trick. Having taken this trick, the world-sheet  fields should be
extended to the whole complex plane. Thus we must consider the following change of variables

\begin{displaymath}
\tilde{X}^{\mu}(\bar{z}) \rightarrow D^{\mu}_{\nu}X^{\nu}(\bar{z}) \ ,
\spa
\tilde{\psi}^{\mu}(\bar{z}) \rightarrow
D^{\mu}_{\nu}\psi^{\nu}(\bar{z}) \ ,
\spa
\tilde{\phi}(\bar{z}) \rightarrow \phi(\bar{z})\,, \mand
\tilde{S}_{\al}(\bar{z}) \rightarrow M_{\al}{}^{\be}{S}_{\be}(\bar{z})
 \ ,
\non\end{displaymath}

Notice that in the last relation left-moving spin operator gets replaced with the product of the constant M matrix and  complex spin operator.
The definitions of $D$ and $M$ matrix are :
\begin{displaymath}
D = \left( \begin{array}{cc}
-1_{9-p} & 0 \\
0 & 1_{p+1}
\end{array}
\right)  \mand
M_p = \left\{\begin{array}{cc}\frac{\pm i}{(p+1)!}\ga^{a_{1}}\ga^{a_{2}}\ldots \ga^{a_{p+1}}
\eps_{a_{1}\ldots a_{p+1}}\,\,\,\,{\rm for\, p \,even}\\ \frac{\pm 1}{(p+1)!}\ga^{a_{1}}\ga^{a_{2}}\ldots \ga^{a_{p+1}}\ga_{11}
\eps_{a_{1}\ldots a_{p+1}} \,\,\,\,{\rm for\, p \,odd}\end{array}\right.
\non\end{displaymath}
Now we are allowed to use just the  holomorphic correlators for all fields  $X^{\mu},\psi^{\mu}, \phi$ as
\begin{eqnarray}
\lan X^{\mu}(z)X^{\nu}(w)\ran & = & -\eta^{\mu\nu}\log(z-w) , \non \\
\lan \psi^{\mu}(z)\psi^{\nu}(w) \ran & = & -\eta^{\mu\nu}(z-w)^{-1} \ ,\non \\
\lan\phi(z)\phi(w)\ran & = & -\log(z-w) .\
\eeqa
The Wick-like rule \cite{Garousi:2007fk,Liu:2001qa}
and \cite{Kostelecky:1986xg} has been used to find out the correlation function between two spin operators and several fermion fields such that

$\frac{}{}$

 \beqa
 \lan\psi^{\mu_1}
(x_1)...
\psi^{\mu_n}(x_n)S_{\al}(z)S_{\be}(\bz)\ran&\!\!\!\!=\!\!\!\!
&\frac{1}{2^{n/2}}
\frac{(z-\bz)^{n/2-5/4}}
{|x_1-z|...|x_n-z|}\left[(\Gamma^{\mu_n...\mu_1}
C^{-1})_{\al\be}\right.\nonumber\\&&+
\lan\lan\psi^{\mu_1}(x_1)\psi^{\mu_2}(x_2)\ran\ran(\Gamma^{\mu_n...\mu_3}
C^{-1})_{\al\be}
\pm perms\nonumber\\&&+\lan\lan\psi^{\mu_1}(x_1)\psi^{\mu_2}(x_2)\ran\ran
\lan\lan\psi^{\mu_3}(x_3)\psi^{\mu_4}(x_4)\ran\ran(\Gamma^{\mu_n...\mu_5}
C^{-1})_{\al\be}\nonumber\\&&\left.
\pm {\rm perms}+\cdots\right],\labell{wicklike}\eeqa
Note that
the  summation  on all possible contractions must be assumed. $\Gamma^{\mu_{n}...\mu_{1}}$ has to be a total antisymmetric matrix in terms  of the gamma matrices. The Wick-like contraction is expressed  as
\beqa
\lan\lan\psi^{\mu}(x_{1})\psi^{\nu}(x_{2})\ran\ran &=&\eta^{\mu\nu}{\frac {(x_{1}-z)(x_{2}-\bz)+(x_{2}-z)(x_{1}-\bz)}{(x_{1}-x_{2})(z-\bz)}}\nonumber\\
&=&2\eta^{\mu\nu}{\frac {Re[(x_{1}-z)(x_{2}-\bz)]}{(x_{1}-x_{2})(z-\bz)}},
\eeqa
Notice the fact that  $x_1,\,x_2$ must be real.
Applying \reef{wicklike} we can easily get the correlation function between two spin operators and one fermion field as

\beqa
I_5^c&=&<:S_{\al}(x_4):S_{\be}(x_5):\psi^c(x_3):>=2^{-1/2}x_{45}^{-3/4}(x_{34}x_{35})^{-1/2}
(\gamma^{c}C^{-1})_{\alpha\beta}.\label{12}
\eeqa
where $\Ga^{\mu\nu}=(\ga^{\mu}\ga^{\nu}-\ga^{\nu}\ga^{\mu})/2$ and we have
defined $x_4\equiv z=x+iy$ ,  $x_5\equiv
\bar{z}$ and $x_{ij} =x_i-x_j$.

\vskip 0.1in

Truly we were able to extend  the Wick-like rule such that  the correlation function of two spin operators and a number of mixed fermions and currents can be achieved \cite{Hatefi:2010ik,Garousi:2008ge}, provided the fact that one must remove Wick-like rule for  two fermion fields $\psi$s in one current.
 Considering this point, we  found consistent results in the absence and presence of current  as follows

\beqa  {<:S_{\al}(x_4):S_{\be}(x_5):>}&=&x_{45}^{-5/4}C^{-1}_{\al\be}, \labell{exp22}\\
<:S_{\al}(x_4):S_{\be}(x_5):\psi^e\psi^f(x_1):>&=& -\frac{1}{2}x_{45}^{-1/4}
x_{14}^{-1}x_{15}^{-1}(\Gamma^{ef}C^{-1})_{\al\be}.\nonumber\eeqa

In the second formula of \reef{exp22}, we just removed Wick-like rule between two fermions in $x_1$ place and got consistent results .

\section{Appendix B:
Some useful integrals for five-point \\
 functions}

 To get some ideas let us mention the method that we have solved  the integrals for five-point functions.
We have gauge fixed by fixing the position of three open string operators and we are left with
  double integrals on $z,\bar z$. They are related to the RR closed string. Thus after gauge fixing we get the following double integrals:

\beqa
I = \int_{\cal{H}^{+}} d^2 \!z |1-z|^{a} |z|^{b} (z - \bar{z})^{c}
(z + \bar{z})^{d},
\nonumber\eeqa
where $d=0,1,2$ and $a,b,c$ should be computed in terms of the Mandelstam variables. Since we are talking about disk level amplitude
the integrations must be done on upper half plane. The necessary conditions for these integrals must be taken into account as
\beqa
a+b+c \leq -2    \nonumber \\
a+b+d \leq -2
\nonumber\eeqa
To remove integrals on $x,y$ we may use the following definitions

\beqa
|z|^{b}= \frac{1}{\Gamma(- \frac{b}{2})} \int_{0}^{\infty} du \,u^{-\frac{b}{2} - 1} e^{-u |z|^2}, \nonumber\\
|1-z|^{a}= \frac{1}{\Gamma(- \frac{a}{2})} \int_{0}^{\infty}
ds\,s^{-\frac{a}{2} - 1} e^{-s |1-z|^2}. \nonumber
\nonumber\eeqa
where
 $z=x+iy$

\beqa
I_{y}= \int^{\infty}_{0} dy \; y^c e^{-(s+u)y^2} =
\frac{\Gamma(\frac{1+c}{2})}{2(s+u)^{\frac{1+c}{2}}},\nonumber
\nonumber\eeqa
\begin{eqnarray}
&F(\lambda)= \displaystyle \int_{-\infty}^{\infty} dx e^{-(s+u)x^2 + 2 \lambda x}  = \sqrt{\frac{\pi}{s+u}} e^{\frac{\lambda^2}{s+u}}.\nonumber
\end{eqnarray}
Here we rewrite down the integration on
$x$

\begin{eqnarray}
I_x= 2^d \int^{\infty}_{-\infty} dx \; x^d e^{-s}e^{-(s+u)x^2+2sx}
= 2^d e^{-s}\int^{\infty}_{-\infty} dx \; x^de^{-(s+u)(x-\frac{s}{s+u})^2+\frac{s^2}{s+u}}.
\end{eqnarray}

so after all the integration on $x$ will be appeared as
\begin{eqnarray}
I_x= 2^d e^{-\frac{us}{s+u}}\int^{\infty}_{-\infty} dx \; x^d
e^{-(s+u)(x-\frac{s}{s+u})^2}= e^{-s} \frac{d \;}{d^d \! \lambda}
F(\lambda)|_{\lambda=s}.
\end{eqnarray}
 $d=n, \ \ n \in Z $,
$$
\label{xInteg} I_{x} = 2^{d} e^{- \frac{us}{s+u}}
\frac{\sqrt{\pi}}{(s+u)^{\frac{1}{2}}} \left \{
\begin{array}{cc}
1 & , d=0 \\
\frac{s}{(s+u)} &, d=1
\end{array} \right.
$$

For simplicity we just do the integration for
$d=0$
and finally we show our results for
$d=1,2$. So for $d=0$ after replacing those steps mentioned above and doing the integrals over $x,y$, collecting them and replacing in the general  integration on $I$,
we will have

\beqa
I= \frac{\sqrt{\pi}(2i)^{c+1} 2^{0}\Gamma(\frac{1+c}{2})}
{2\Gamma(\frac{-a}{2})
\Gamma(\frac{-b}{2})}
\int^{\infty}_{0}\int^{\infty}_{0}
\frac{ds du}{(s+u)^{1+c/2}} u^{-b/2-1}s^{-a/2-1}e^{- \frac{us}{s+u}},
\eeqa
We might use the following change of variables
\beqa
s=\frac{x}{t},\quad u=\frac{x}{1-t},\quad dsdu=Jdxdt=\frac{xdxdt}{(t(1-t))^2}
\nonumber\eeqa

Replacing the change of variables in the Jacobian, we find
\beqa
I= \frac{\pi^{1/2}(2i)^{c+1}\Gamma(\frac{1+c}{2})}{2\Gamma(\frac{-a}{2})\Gamma(\frac{-b}{2})}
\int^{\infty}_{0}dxe^{-x}x^{\frac{-4-(a+b+c)}{2}}\int^{1}_{0}dt  t^{(c+a)/2}(1-t)^{(c+b)/2},
\eeqa

\beqa
I = (2 \imath)^{c}  \,  \pi \frac{ \Gamma( 1 +
\frac{b+c}{2})\Gamma( 1+ \frac{a+c}{2})\Gamma( -1-
\frac{a+b+c}{2})\Gamma( \frac{1+c}{2})}{
\Gamma(-\frac{a}{2})\Gamma(-\frac{b}{2})\Gamma(2+c+
\frac{a+b}{2})}.
\nonumber\eeqa

The following relations have been used
\beqa
\int^{1}_{0}dx  x^{\beta-1}(1-x)^{\alpha-1}=\frac{\Gamma(\alpha)\Gamma(\beta)}{\Gamma(\alpha+\beta)},\quad \Gamma(z)=(z-1)!,
\nonumber
\eeqa
Eventually we obtain the result for
$d=1$
as

\beqa
I = (2 \imath)^{c} 2 \,  \pi \frac{ \Gamma( 2 +
\frac{b+c}{2})\Gamma( 1+ \frac{a+c}{2})\Gamma( -1-
\frac{a+b+c}{2})\Gamma( \frac{1+c}{2})}{
\Gamma(-\frac{a}{2})\Gamma(-\frac{b}{2})\Gamma(3+c+
\frac{a+b}{2})},\nonumber
\eeqa
Therefore one  can write them down in a closed form as
\beqa
 \int d^2 \!z |1-z|^{a} |z|^{b} (z - \bar{z})^{c}
(z + \bar{z})^{d}&\!\!\!\!=\!\!\!&
(2i)^{c} 2^d \,  \pi \frac{ \Gamma( 1+ d +
\frac{b+c}{2})\Gamma( 1+ \frac{a+c}{2})\Gamma( -1-
\frac{a+b+c}{2})\Gamma( \frac{1+c}{2})}{
\Gamma(-\frac{a}{2})\Gamma(-\frac{b}{2})\Gamma(2+c+d+
\frac{a+b}{2})}.\nonumber
\eeqa
 The above result is valid For $d=0,1$  \cite{Fotopoulos:2001pt}. Applying the same method for $ d= 2$, one sets \cite{Garousi:2008ge}
\beqa
 \int d^2 \!z |1-z|^{a} |z|^{b} (z - \bar{z})^{c}
(z + \bar{z})^{d}&\!\!\!\!=\!\!\!&
(2i)^{c} 2^d \,  \pi \frac{J_1+J_2}{
\Gamma(-\frac{a}{2})\Gamma(-\frac{b}{2})\Gamma(d+2+c+
\frac{a+b}{2})}.\eeqa
where
\beqa
J_1&=&\frac{1}{2}\Gamma( d+
\frac{b+c}{2})\Gamma( d+ \frac{a+c}{2})\Gamma( -d-
\frac{a+b+c}{2})\Gamma( \frac{1+c}{2})\nonumber\\
J_2&=&\Gamma( d+1+
\frac{b+c}{2})\Gamma( 1+ \frac{a+c}{2})\Gamma( -1-
\frac{a+b+c}{2})\Gamma( \frac{1+c}{2}).\nonumber\eeqa



\end{document}